\documentclass[reprint,superscriptaddress,aip,jcp,]{revtex4-2}
\usepackage{graphicx}
\usepackage{placeins}
\usepackage{color}
\usepackage{latexsym}
\usepackage{amsmath}
\usepackage{amsfonts}
\usepackage{amssymb}
\usepackage{bm}
\usepackage{graphicx}
\usepackage{mathtools}
\usepackage{amsbsy}
\usepackage{physics}
\usepackage{enumitem}
\usepackage{lipsum}
\usepackage{rotating}
\newcommand{\dt}{\Delta t}
\expandafter\ifx\csname package@font\endcsname\relax\else
 \expandafter\expandafter
 \expandafter\usepackage
 \expandafter\expandafter
 \expandafter{\csname package@font\endcsname}%
\fi
\usepackage[normalem]{ulem}

\graphicspath{{./figures/},{figures}}

\newcommand{\lla}{\left\langle}
\newcommand{\rra}{\right\rangle}

\newcommand{\psum}{\sideset{}{'}\sum}

\let\Gamma\varGamma
\let\Delta\varDelta
\let\Theta\varTheta
\let\Lambda\varLambda
\let\Pi\varPi
\let\Sigma\varSigma
\let\Upsilon\varUpsilon
\let\Phi\varPhi
\let\Psi\varPsi
\let\Omega\varOmega
\let\phi\varphi

\newcommand{\cred}{\color{black}}

\newcommand{\REV}[1]{\textcolor{black}{#1}}

\usepackage{hyperref}

\begin{document}
\title{Virial Stress in Systems of Active Brownian Particles in the Presence of Translational and Rotational Inertia}

\author{Chandranshu Tiwari}%
\email{chandranshu21@iiserb.ac.in}
\affiliation{Department of Physics, Indian Institute of Science Education and Research, \\ Bhopal 462 066, Madhya Pradesh, India}
\author{Sunil P. Singh}
 \email{spsingh@iiserb.ac.in}
\affiliation{Department of Physics, Indian Institute of Science Education and Research, \\ Bhopal 462 066, Madhya Pradesh, India}
\author{Roland G. Winkler}%
\email{rg\_winkler@gmx.de}
\affiliation{Theoretical Physics of Living Matter, Institute for Advanced Simulation, Forschungszentrum J{\"u}lich, 52425 J{\"u}lich, Germany}

\date{\today}

\begin{abstract}
We elucidate the stress in a system of active Brownian particles augmented with translational and rotational inertia (ABP+TRI). Stress tensors are derived for periodic systems as well as systems confined between walls by employing Lagrange's equations of motion of the first kind for the rotational motion. \REV{Using Langevin simulations of an ideal active gas in two dimensions, we confirm the existence of an equation of state for periodic systems that depends on translational and rotational inertia in general.} Confinement implies a strong polarization of the propulsion direction near a wall and an enhanced density, both of which increase with increasing rotational inertia. This affects the local stress tensor normal to the confining walls, leading to a breakdown of the equation of state. Yet the local stress in the bulk part of the confined systems is identical with that of the periodic system. Importantly, for both kinds of boundary conditions, the so-called swim stress is not included in the local stress tensor; thus, in general, the swim stress is not representative of the stress in systems of ABP+TRIs.   
\end{abstract}

\maketitle

\section{Introduction}

Active matter, whose agents consume internal energy or extract energy from the
environment, constitutes an outstanding class of nonequilibrium systems.~\cite{rama:10,marc:13,cate:15,wink:15,elge:15,bech:16,shae:20,gomp:20,omar:20} The active-matter spectrum ranges from microscopic entities, such as bacteria and algae, to macroscopic systems, as flocks of birds, schools of fish, granular matter, human crowds, and autonomous robots.~\cite{rama:10,marc:13,cate:15,wink:15,elge:15,bech:16,shae:20,gomp:20} So far, a comprehensive statistical description based on active units is lacking. Yet, particle-based computer simulations~\cite{elge:15,bech:16,shae:20} of active systems reveal fascinating features such as motility-induced phase separation (MIPS),~\cite{bial:12,fily:12,wyso:14,cate:15,bech:16} surface accumulation,~\cite{tiwari2024collective,
yang:14.2,fily:14,elge:13.1,solo:15.1,das:19} or active turbulence.~\cite{wens:12,qi:22,zant:22} Especially, mechanical properties are accessible by such simulations, and provide insight into implications of active motion onto thermodynamic properties.~\cite{taka:14,fily:14,taka:15,wink:15,solo:15.1,fily:18} In particular, the presence of an equation of state has been intensively studied.~\cite{solo:15.1,spec:16,wink:15,fily:18,das:19,omar:20,sand:23} 

The existence of a state equation depends on a suitable definition of the stress (pressure) in an active system, in particular, the definition of a local stress (pressure).~\cite{das:19,fily:18} In certain systems, specifically systems with periodic boundary conditions, the active stress, and a resulting equation of state, is correctly described by the so-called swim stress~\cite{taka:14}---the virial by the active force.~\cite{taka:14,wink:15,das:19} However, this term does not provide the correct local stress equation-of-state in the presence of confining walls, spatial inhomogeneities,~\cite{das:19,fily:18} or orientational-dependent active forces.~\cite{solo:15.1,fily:18} 

The presence of translational and rotational inertia of active particles renders the derivation of a local stress even more demanding, as spatial inhomogeneities, e.g., by confining walls, induce additional stresses. Simulations of systems consisting of active Brownian particles with translational and rotational inertia (ABP+TRI) reveal a pronounced influence of inertia on their phase behavior. In particular, translational inertia suppresses MIPS,~\cite{loew:20} whereas an increasing rotational inertia drastically enhances the stability of the phase coexistence.~\cite{capr:22.1}

To gain deep insight into such phenomena, it is crucial to derive a stress tensor for systems of underdamped active particles. Specifically, the contributions by translational and rotational inertia need to be accounted for. In general, an accurate formulation of the stress tensor is essential for investigating mechanical stability and thermodynamic properties of active matter systems.

In this article, we systematically derive stress tensors for systems of active Brownian particles with translational and rotational inertia from their translational and rotational equations of motion. For the latter, we use Lagrange's equations of the first kind~\cite{gold:80} extended by a stochastic force. We derive virial expressions for both the global and local stress tensors in systems with periodic boundary conditions and in systems confined between walls.

For periodic systems, the global as well as local stress consists of the well-known kinetic stress, an additional contribution from the swim momentum,~\cite{das:19} and a particular term depending on the moment of inertia, denoted as angular-velocity stress in the following. We demonstrate that the local and global stresses are identical in a gas of ABP+TRIs, indicating the existence of an equation of state. Importantly, in the presence of inertia, the swim stress~\cite{taka:14,das:19} does not characterize the active contribution to the overall stress anymore. Hence, the sum of kinetic and swim stress does not provide an equation of state, as it does in the case of a zero moment of inertia.~\cite{das:19} Remarkably, there is a tight coupling between the translational and rotational inertia. For small translational inertia (small mass $m$), the stress tensor is essentially independent of the moment of inertia. On the contrary, for large $m$, the moment of inertia changes the stress considerably.       

The presence of confining walls substantially affects the stress in an ABP+TRI system. It leads to an enhanced particle density adjacent to a wall and, more importantly, to a pronounced polarization of the ABP+TRI propulsion direction toward the wall.~\cite{yan:15,ezhi:15,elge:15,fily:18,ausc:21,carr:24} These effects imply an increased magnitude of the local stress that decreases slowly toward the bulk. Hence, in combination with the wall, activity acts as a local momentum source.~\cite{fily:18} As a consequence, there is no equation of state in such a system.~\cite{fily:18} We explicitly calculate the stress as force per area on a wall and compare it with the internal stress, which accounts for all particles within the considered volume. The equivalence of the two expressions does not allow any conclusion to be drawn about the existence of an equation of state, as the internal stress does not necessarily correspond to a local stress or a bulk stress. Our derived expression for the local stress includes additional terms that account for a possible polarization of the propulsion direction. 
While rotational inertia increases persistence, thereby enhancing surface accumulation and polarization, translational inertia counteracts this effect by causing particles to reflect from a wall. As a result, systems with large translational inertia generally exhibit reduced surface accumulation and polarization. Notably, in the limit of vanishing moment of inertia, the bulk-polarization term disappears, and the stress tensor for the confined and periodic system becomes identical for a vanishing range of surface interaction. Hence, in this limit, an equation of state exists, as already discussed in Refs.~\onlinecite{das:19,fily:18}. Moreover, even in the presence of inertia, the stress in the bulk part is equal to that in a periodic system at the same bulk density.    
\section{Model}

We consider a system of $N$ self-propelled point-like particles with translational and rotational inertia in $d$ dimensions ($d=2,3$). The translational dynamics of the $i^{th}$ particle ($i \in \{1,\ldots,N\}$) of mass $m$ and position ${\bm r}_i(t)$, $t$ being the time, is described by the Langevin equation
\begin{equation}\label{eq:eqm_r}
    m \ddot{\bm r}_i(t) =- \gamma \dot{\bm r}_i(t) + \gamma \bm v^{(a)}_i(t) + \bm F_i(t) + \bm \varGamma_i(t) ,
\end{equation}
where $\dot{\bm r}_i$ is the velocity, $\gamma$ is the translational friction coefficient, and $\bm v_i^{(a)} = v_0 \bm e_i$ is the propulsion velocity with the active speed $v_0$ in the direction $\bm e_i$. The conservative forces $\bm F_i$ include pairwise interparticle interactions, $\bm F_{ij}$, as well as possible forces by the presence of confining walls, $\bm F_i^{(w)}$. \cite{wink:15,das:19} Thermal fluctuations are captured by the stochastic force ${\bm \varGamma}_{i}$, a Gaussian and Markovian stochastic process of zero mean, and the correlations
\begin{equation}
  \lla  \varGamma_{\alpha i}^T(t) \varGamma_{\beta j}^T(t^{\prime}) \rra = 2 \gamma k_BT {\delta}_{\alpha \beta} {\delta}_{i j} \delta(t-t^{\prime}) ,   
\end{equation}
with $\alpha, \beta \in \{x,y,z\}$, $T$ the temperature, and $k_B$ the Boltzmann constant. 
 
Present excluded-volume interactions between ABP+TRIs are described by the truncated, repulsive Lennard-Jones potential
\begin{equation}
U_{LJ} = 4\epsilon \sum_{i = 1}^{N - 1}\sum_{j = i+1}^{N}  \left[\left(\frac{\sigma}{r_{ij}}\right)^{12} - \left(\frac{\sigma}{r_{ij}}\right)^{6}  + \frac{1}{4} \right], 
\label{Eq:LJ}
\end{equation}
for $r_{ij} \le  2^{1/6}\sigma$ and zero otherwise. In Eq.~\eqref{Eq:LJ}, $r_{ij} = |\bm{r}_{i} - \bm{r}_{j}|$ is the distance between a particle pair $i$ and $j$, $\epsilon$ the strength of the potential, and $\sigma$ the range of the interaction. Particle-wall interactions are described by a similar Lennard-Jones potential.
 
We express the dynamics of a particle's propulsion direction in Cartesian coordinates. This allows us to capture the tight coupling between the translational and rotational motion of an ABP+TRI in deriving a stress tensor. Considering the propulsion directions as vectors obeying the constraints $|\bm e^2_i(t)| =1$, their equations of motion are given by the Langevin equations 
 \begin{equation} \label{eq:eqm_e}
    I \ddot{\bm e}_i(t) = - \gamma_r \dot{\bm e}_i(t) + \lambda_i \bm e_i(t) +  \gamma_r \left(\bm \varGamma_i^e(t) \times \bm e_i(t) \right).  
\end{equation}
The stochastic forces  $\bm \varGamma_i^e$ are also assumed as Gaussian white noise with zero mean and the correlations 
\begin{equation}
  \lla  \varGamma_{\alpha i}^e(t) \varGamma_{\beta j}^e(t^{\prime}) \rra = 2 D_r {\delta}_{\alpha \beta} {\delta}_{ij} \delta(t-t^{\prime}) ,   
\end{equation}
with the rotational diffusion coefficient $D_r$. In the following, we will use the abbreviation $\gamma_R = (d-1) D_r$.~\cite{wink:15,das:19} 

\REV{The equation of motion, Eq.~\eqref{eq:eqm_e}, equivalently describes the rotational dynamics of a solid body---rod or disc---in two dimensions. Yet, in three dimensions, it corresponds to the rotational dynamics of an infinitely thin rod rather than a sphere, as Eq.~\eqref{eq:eqm_e} implies that the axis for the solid-body-type rotational dynamics is orthogonal to the propulsion direction. Qualitatively, this yields the same rotational dynamics in three dimensions. Even the correlation function for the propulsion direction $\bm e(t)$ decays similarly to the prediction in Ref.~\onlinecite{sand:23} for short times. Merely, the long-time behavior differs, as preliminary simulations show.} 

The translational equations of motion are solved by the integration scheme presented in Ref.~\onlinecite{gron:13}. For the integration of the equations of motion of the propulsion directions, we use a tailored version of the RATTLE algorithm (cf. App.~\ref{app:integ_e}).~\cite{ande:83}

{\it Parameters:} We use $\sigma$ as unit of length, $k_BT$ of energy, and $\tau_r = 1/D_r$ of time. The propulsion strength is characterized by the P\'eclet number 
 \begin{equation}
     Pe = \frac{v_0}{\sigma D_r}.
 \end{equation} 
The ratio between the translational diffusion coefficient of a passive particle, $D_0=k_BT/\gamma$, and the rotational diffusion coefficient is fixed by setting $D_0 = k_BT/\gamma = \sigma^2 D_r$. The dimensionless mass and moment of inertia are given by
 \begin{align}
 M = \frac{ m D_r}{\gamma} = \frac{\tau_m}{\tau_r} , \\
 J = \frac{ I D_r}{\gamma_r} = \frac{\tau_I}{\tau_r} ,  
 \end{align}
where $\tau_m = m/\gamma$ and $\tau_I =I/\gamma_r$ are translational and rotational time scales.~\cite{loew:20} The equations of motion are integrated with the time step $\Delta t = 10^{-3} \tau_r$. 

In the simulations, two-dimensional systems with square simulation boxes of the side length $L$ are considered. Either periodic boundary conditions are applied, or confinement by two walls normal to the $x$ axis and periodic boundary conditions along the $y$ direction. 
In case of periodic boundary conditions, we use $L/\sigma = 200$ and the number of particles $N=12732$, which corresponds to the number density $\bar \rho = N / L^2 = 0.318/\sigma^2$. In case of confinement,  we consider $L/\sigma =1000$ and $N=318,309$,
while maintaining the same number density as for the periodic system.  

\section{Stress Tensor of Confined System}

\subsection{Global stress tensor} \label{sec:global_conf}

The stress tensor $\bm \sigma$ of a mechanical system can be derived via Clausius' virial theorem.~\cite{beck:67,wink:92.1,wink:93,wink:15} In this derivation, the Cartesian components $\alpha$ of the equations of motion Eq.~\eqref{eq:eqm_r} are multiplied by $r_{\alpha i}$.~\cite{wink:92.1,wink:93,wink:15,das:19}
Averaging (time or ensemble average) and summation over all particles yields then
\begin{align} \label{eq:virial_tot_conf} \nonumber
 m \sum_i \lla \dot{r}^2_i \rra  &  + \gamma \sum_i \lla v_i^{(a)} r_i  \rra + \frac{1}{2} \sum_{i=1}^N \psum_{j=1}^N \lla F_{ij}r_{ij} \rra  \\ & + \sum_{i=1}^N \lla F^w_{i}r_{i} \rra   =0 ,
\end{align} 
with the identities $d(\dot{\bm r}_i \cdot \bm r_i)/dt = \ddot{\bm r}_i \cdot \bm r_i + \dot{\bm r}_i^2$, $d \bm r^2_i/dt = 2 \bm r_i \cdot \dot{\bm r}_i$, and $d \langle r_{\alpha i} \dot{r}_{\alpha i} \rangle/dt  = d \langle r_{\alpha i}^2\rangle/dt =0$ due to confinement along the Cartesian axis $\alpha$.~\cite{wink:15} (Here, and in some of the subsequent expressions, the index $\alpha$ is suppressed.)  The average over the stochastic forces vanishes in the presence of inertia. The prime in the double sum indicates that the index $j = i$ is excluded.

In a confined system, with walls located at $\pm L_{\alpha}/2$, we can introduce an external stress component along the axis $\alpha$, $\sigma_{\alpha \alpha}^{(e)}$, as the average total force per area exerted on the confining walls and an internal global virial stress, $\sigma_{\alpha \alpha}^{(i)}$, as \cite{wink:93,wink:15,das:19}
\begin{widetext}
\begin{align}  \label{eq:stress_ext_conf}
V \sigma_{\alpha \alpha}^{(e)} = & \   \sum_{i=1}^N \lla F_{\alpha i}^{(w)} \rra L_{\alpha i},  \\ \label{eq:stress_int_conf}
V \sigma_{\alpha \alpha}^{(i)} =  & \ - m \sum_{i=1}^N  \lla \dot{r}_{\alpha i}^2 \rra - \gamma  \sum_{i=1}^N  \lla v_{\alpha i}^{(a)} r_{\alpha i} \rra  - \sum_{i=1}^N \lla F^{(w)}_{\alpha i}(r_{\alpha i}-L_{\alpha i}) \rra - \frac{1}{2} \sum_{i=1}^N \psum_{j=1}^N \lla F_{\alpha ij}r_{\alpha ij}  \rra ;
\end{align}
\end{widetext}
$V$ is the total volume. Here, $L_{\alpha i} = \pm L_{\alpha}/2$ is the Cartesian coordinate of the wall, with which the $i^{th}$ particle interacts.  
The wall term in Eq.~\eqref{eq:stress_int_conf} accounts for a possible finite range of the wall forces. The contribution vanishes in the limit of zero-range wall forces. The external and internal stress tensors are equal by definition (Eq.~\eqref{eq:virial_tot_conf}). In an isotropic system, the pressure is given by $p=-\sum_{\alpha} \sigma_{\alpha \alpha}/d$.

The internal stress tensor in Eq.~\eqref{eq:stress_int_conf} includes kinetic contributions from inertia
\begin{equation} \label{eq:kin_stress}
  V \sigma_{\alpha \alpha}^{(k)} =   - m \sum_{i=1}^N  \lla \dot{r}_{\alpha i}^2 \rra ,
\end{equation}
which comprises thermal fluctuations as well as contributions from the activity of the particles, the so-called swim stress  \cite{taka:14,wink:15,smal:15,das:19}
\begin{equation} \label{eq:swim_stress}
   V \sigma_{\alpha \alpha}^{(s)} = - \gamma  \sum_{i=1}^N  \lla v_{\alpha i}^{(a)} r_{\alpha i} \rra ,
\end{equation}
as well as contributions by inter-particle forces (last term). 

The external stress Eq.~\eqref{eq:stress_ext_conf} is the mechanical force exerted on a wall, and the internal stress Eq.~\eqref{eq:stress_int_conf} is the equivalent expression comprising contributions of all particles in the volume along the spatial direction ${\alpha}$. However, this does not necessarily prove the existence of an equation of state, as the stress in the bulk may differ from that at the wall,~\cite{solo:15.1} as we will demonstrate in Sec.~\ref{sec:confined_sim}. As a consequence, the swim stress does not represent the active stress in general.~\cite{fily:18,das:19}

\subsection{Local stress tensor}

To derive a local stress tensor at the position $r_{\alpha}$ between confining walls, we consider the interval $\Delta L_{\alpha}$ centered around $r_{\alpha}$, where the total subvolume of a cuboid is $\Delta V = \Delta L_x \times \Delta L_y \times \Delta L_z$.\cite{das:19} 
Active particles inside of $\Delta V$ interact with each other as well as particles outside. In addition, particles enter and leave $\Delta V$ in the course of time. The virial stress of particles in the interval $\Delta L_{\alpha}$ can be obtained by multiplying  Eqs.~\eqref{eq:eqm_r} and \eqref{eq:eqm_e} by $r_{\alpha i}$, an additional factor $\Lambda_{i}(\bm r)$, and by averaging and summing over all particles. The factor $\Lambda_{i}(\bm r)$ accounts for the intervals $\Delta L_{\alpha}$---$\Lambda_{i}(\bm r)$ is unity when particle $i$ is within $\Delta V$ and zero otherwise.~\cite{das:19,lion:12} As a result, we obtain
\begin{widetext}
\begin{align} \label{eq:virial_r}
m \sum_{i=1}^N \lla \dot{r}_{i}  r_{i} \dot \Lambda_{i} \rra + \frac{\gamma}{2} \sum_{i=1}^N \lla r_{i}^2 \dot \Lambda_{i} \rra  + m \sum_{i=1}^N \lla \dot{r}_i^2 \Lambda_{i} \rra  +  \sum_{i=1}^N \lla F_{i} r_{i} \Lambda_{i} \rra + \gamma \sum_{i=1}^N  \lla  v^{(a)}_{i} r_{i} \Lambda_{i} \rra = & \ 0 ,\\  \label{eq:virial_e}
I \sum_{i=1}^N  \lla  \dot e_{i} r_{i} \dot \Lambda_{i} \rra 
+  \gamma_r \sum_{i=1}^N  \lla  e_{i} r_{i} \dot \Lambda_{i} \rra +  I \sum_{i=1}^N  \lla  \dot{e}_{i} \dot r_{i} \Lambda_{i} \rra  + \gamma_r \sum_{i=1}^N  \lla  e_{i} \dot r_{i} \Lambda_{i} \rra +   \sum_{i=1}^N  \lla \lambda_i  e_{i} r_{i} \Lambda_{i} \rra    = & \ 0 ,
\end{align}
\end{widetext}
with the identities $\langle d(\dot{\bm r}_i \cdot \bm r_i \Lambda_i)/dt \rangle = \langle \ddot{\bm r}_i \cdot \bm r_i \Lambda_i \rangle  + \langle \dot{\bm r}_i^2 \Lambda_i \rangle + \langle \dot{\bm r}_i \cdot \bm r_i \dot \Lambda_i \rangle =0$ , $\langle d( \bm r^2_i \Lambda_i)/dt \rangle = \langle 2 \bm r_i \cdot \dot{\bm r}_i \Lambda_i \rangle + \langle \bm r^2_i \dot \Lambda_i \rangle =0$, $\langle d(\dot{\bm e}_i \cdot \bm r_i \Lambda_i)/dt \rangle = \langle \ddot{\bm e}_i \cdot \bm r_i \Lambda_i \rangle  + \langle \dot{\bm e}_i \cdot \dot{\bm r}_i \Lambda_i \rangle  + \langle \dot{\bm e}_i \cdot {\bm r}_i \dot \Lambda_i \rangle =0$, and $\langle d({\bm e}_i \cdot \bm r_i \Lambda_i)/dt \rangle = \langle \dot{\bm e}_i \cdot \bm r_i \Lambda_i \rangle  + \langle {\bm e}_i \cdot \dot{\bm r}_i \Lambda_i \rangle  + \langle {\bm e}_i \cdot {\bm r}_i \dot \Lambda_i \rangle =0$  due to confinement. 
The averages over terms with stochastic forces are zero under the Ito interpretation of stochastic processes. The terms with $\dot \Lambda_i$ account for the exchange of particles between the volume $\Delta V$ and its surrounding.~\cite{lion:12}  The other terms in Eq.~\eqref{eq:virial_r} are similar to those in Eq.~\eqref{eq:virial_tot_conf}. Here, we assume that there are no external bulk forces.  

In Sec.~\ref{sec:global_conf}, the stress tensors have solely been derived using the translational equations of motion and virials of the particles (Eq.~\eqref{eq:virial_tot_conf}). Then, the active term leads to the swim stress. Considering the last term of Eq.~\eqref{eq:virial_r} of a confined system, this term is locally zero in an isotropic and homogeneous system, as $r_{\alpha i} = r_{\alpha}=const.$ as a function of time and the average of $\langle e_{\alpha i} \rangle$ is zero. This would suggest that there is locally no explicit contribution of the propulsion force to stress.~\cite{spec:16} However, the virial equation Eq.~\eqref{eq:virial_r} is not the defining equation for a stress tensor, as already shown in Ref.~\onlinecite{das:19} for an overdamped rotational motion. To derive a local stress tensor, we  multiply  Eq.~\eqref{eq:virial_e} by $v_0\gamma/(\gamma_r \gamma_R) = v_0\gamma/((d-1) \gamma_r D_r)$, add and subtract the term  $\gamma \sum_i  \langle v_i^{(a)} r_i\Lambda_i\rangle$, and replacing this term in Eq.~\eqref{eq:virial_r}. This leads to the equation
\begin{widetext}
\begin{align}  \nonumber \label{eq:virial_local_tot}
& m \sum_{i=1}^N  \lla \dot{r}_{i}  r_{i} \dot \Lambda_{i} \rra  + \frac{\gamma}{2} \sum_{i=1}^N \lla {r}^2_{i} \dot \Lambda_{i} \rra   + \frac{\gamma}{\gamma_R} \sum_{i=1}^N  \lla  v^{(a)}_{i} r_{i} \dot \Lambda_{i} \rra + \frac{\gamma I}{\gamma_r \gamma_R} \sum_{i=1}^N  \lla  \dot v^{(a)}_{i} r_{i} \dot \Lambda_{i} \rra\\ & \ + 
m \sum_{i=1}^N \lla \dot{r}_i^2 \Lambda_{i} \rra +  \sum_{i=1}^N \lla F_{i} r_{i} \Lambda_{i} \rra + \frac{\gamma}{\gamma_R} \sum_{i=1}^N  \lla  v^{(a)}_{i} \dot r_{i} \Lambda_{i} \rra + \frac{\gamma I}{\gamma_r \gamma_R} \sum_{i=1}^N  \lla \dot v^{(a)}_{i}  \dot r_{i} \Lambda_{i} \rra  + \frac{\gamma }{\gamma_r \gamma_R}  \sum_{i=1}^N  \lla (\lambda_i + \gamma_r \gamma_R)  v^{(a)}_{i} r_{i}  \Lambda_{i} \rra = 0.
\end{align}
\end{widetext}
The terms with $\dot \Lambda_i$ represent the momentum flux across the boundaries of the volume $\Delta V$, and hence, represent an external stress. In addition, there are external stress contributions by forces between particles inside and outside of $\Delta V$, in analogy to the surface forces in Eq.~\eqref{eq:virial_tot_conf}.\cite{lion:12,das:19} Separating the pairwise forces into contributions from inside and outside of the volume $\Delta V$,  we can define the internal local stress tensor as
\begin{widetext}
\begin{align} \label{eq:local_virial_conf} \nonumber 
 - \Delta V \sigma_{\alpha \alpha}^{(i)}  =  \, & \sum_{i=1}^N m \lla \dot{r}_{\alpha i}^2 \Lambda_{i} \rra + \sum_{i=1}^N \lla F^{(w)}_{\alpha i}(r_{\alpha i}-L_{\alpha i}) \Lambda_i \rra + \frac{1}{2} \sum_{i=1}^N \psum_{j=1}^N \lla F_{\alpha ij} r_{\alpha ij} \eta_{\alpha ij} \rra    \\ & \, + \frac{\gamma}{\gamma_R} \sum_{i=1}^N  \lla  v^{(a)}_{\alpha i} \dot r_{\alpha i} \Lambda_{i} \rra   +  \frac{\gamma I}{\gamma_r \gamma_R} \sum_{i=1}^N  \lla  {\dot v}^{(a)}_{\alpha i} \dot r_{\alpha i} \Lambda_{i} \rra + \frac{\gamma }{\gamma_r \gamma_R}  \sum_{i=1}^N  \lla (\lambda_i + \gamma_r \gamma_R)   v^{(a)}_{\alpha i} r_{\alpha i}  \Lambda_{i} \rra .
\end{align}
\end{widetext}
The factor $\eta_{\alpha ij} =1$, if both particles are inside the volume $\Delta V$---the interparticle term is then reminiscent of the similar term in the global stress tensor. In case of particle $i$ inside and $j$ outside of $\Delta V$ or both are outside, but their connecting line intersects the volume $\Delta V$, then $\eta_{ij}<1$ is the fraction of the distance $|\bm r_i - \bm r_j|$ inside of $\Delta V$. The $\eta_{\alpha ij}$ is the respective Cartesian component of the part of the vector $\bm r_i - \bm r_j$ inside of $\Delta V$ (see Refs.~\onlinecite{das:19,lion:12}). 

An equivalent representation to the global stress tensor in Eq.~\eqref{eq:stress_int_conf} is obtained by choosing the subvolume $\Delta V$ in Eq.~\eqref{eq:local_virial_conf} equal to the total volume $V$, i.e., $\Lambda_i= \eta_{\alpha ij}=1$. This yields the internal stress tensor
\begin{widetext}
\begin{align} \label{eq:virial_conf_int}  \nonumber
 -  V \sigma_{\alpha \alpha}^{(i)}  = & \,  \sum_{i=1}^N m \lla \dot{r}_{\alpha i}^2 \rra + \sum_{i=1}^N \lla F^{(w)}_{\alpha i}(r_{\alpha i}-L_{\alpha i})  \rra  + \frac{1}{2} \sum_{i=1}^N \psum_{j=1}^N \lla F_{\alpha ij} r_{\alpha ij} \rra  \\ & \, +  \frac{\gamma}{\gamma_R} \sum_{i=1}^N  \lla  v^{(a)}_{\alpha i} \dot r_{\alpha i}  \rra  +  \frac{\gamma I}{\gamma_r \gamma_R} \sum_{i=1}^N  \lla  {\dot v}^{(a)}_{\alpha i} \REV{\dot r_{\alpha i}}  \rra + \frac{\gamma }{\gamma_r \gamma_R}  \sum_{i=1}^N  \lla (\lambda_i + \gamma_r \gamma_R)   v^{(a)}_{\alpha i} r_{\alpha i}  \rra .
\end{align}
\end{widetext}

The local stress, Eq.~\eqref{eq:local_virial_conf}, as well as the global stress, Eq.~\eqref{eq:virial_conf_int}, comprise the kinetic stress, $\sigma_{\alpha \alpha}^{(k)}$ (Eq.~\eqref{eq:kin_stress}), and additionally various other terms that are not present in Eq.~\eqref{eq:stress_int_conf}. The stress contribution 
\begin{equation} \label{eq:swim_mom_stress}
    V \sigma_{\alpha \alpha}^{(sm)} = - \frac{\gamma}{\gamma_R} \sum_{i=1}^N \lla v_{\alpha i}^{(a)} \dot r_{\alpha i}  \rra 
\end{equation}
includes the active momentum \cite{das:19,fily:18}
\begin{equation}  \label{eq:active_moment}
\bm p_i^{(a)} = \frac{\gamma}{\gamma_R} \bm v_i^{(a)} , 
\end{equation}
and is therefore referred to as the swim-momentum (sm) stress in the following. In the presence of rotational inertia, the coupling between translational motion and rotational dynamics leads to additional momentum fluxes. As a consequence, the additional stress tensor term 
\begin{equation} \label{eq:ang_velo_stress}
    V \sigma_{\alpha \alpha}^{(av)} = - \frac{\gamma I}{\gamma_r \gamma_R} \sum_{i=1}^N \lla \dot v_{\alpha i}^{(a)} \dot r_{\alpha i}  \rra ,
\end{equation}
appears, with the derivative of the active momentum, i.e., $\dot {\bm e}_i$. Hence, we will denote this term as angular-velocity stress, because $\dot {\bm e}_i = \bm \omega_i \times \bm e_i$, with the angular velocity $\bm \omega_i$. \REV{The angular-velocity stress corresponds to a rotational momentum (angular momentum) flux, since it includes the moment of inertia $I$. Thus, it only appears in presence of rotational inertia.}

The constraint forces (Eq.~\eqref{eq:force_constraint})  imply
the additional stress expression
\begin{equation} \label{eq:const_swim_stress}
V \sigma_{\alpha \alpha}^{(cs)} = - \frac{\gamma }{\gamma_r \gamma_R}  \sum_{i=1}^N  \lla (\lambda_i + \gamma_r \gamma_R)   v^{(a)}_{\alpha i} r_{\alpha i}  \rra ,
\end{equation}
which, aside from a term accounting for the constraint forces, includes the swim stress. We will denote this term as constraint+swim (cs) stress. This term vanishes in various cases, e.g., in the limit $I\to 0$, as will be discussed below (Eq.~\eqref{eq:local_virial_conf_damp}).

The various terms in Eq.~\eqref{eq:virial_conf_int} depend on bulk properties only, aside from the constraint+swim stress. Hence, the latter may yield a wall-dependent contribution to the total stress, which implies the absence of an equation of state.~\cite{fily:18}    

We like to emphasize that even in the absence of rotational inertia, $I=0$, the local swim and swim-momentum stresses differ, because of the terms with the time derivation of $\Lambda_i$ in Eq.~\eqref{eq:virial_e} (see also Ref.~\onlinecite{das:19}). Hence, locally no stress equation and, in particular, no equation of state can be derived solely from Eq.~\eqref{eq:virial_r}.

In the limit $I \to 0$, the Lagrangian multipliers become $\lambda_i = - \gamma_r \gamma_R$ (see App.~\eqref{app:integ_e}, Eq.~\eqref{app:lagpar_limit}), and the two terms in the bracket of the constraint+swim stress term of Eq.~\eqref{eq:local_virial_conf} add up to zero. Then, equation~\eqref{eq:local_virial_conf} becomes 
\begin{align} \label{eq:local_virial_conf_damp} \nonumber 
 - \Delta V \sigma_{\alpha \alpha}^{(i)}  =  \, & \sum_{i=1}^N m \lla \dot{r}_{\alpha i}^2 \Lambda_{i} \rra  + \frac{\gamma}{\gamma_R} \sum_{i=1}^N  \lla  v^{(a)}_{\alpha i} \dot r_{\alpha i} \Lambda_{i} \rra  \\ \nonumber & \ +  \sum_{i=1}^N \lla F^{(w)}_{\alpha i}(r_{\alpha i}-L_{\alpha i})  \rra 
 \\ & \ + \frac{1}{2} \sum_{i=1}^N \psum_{j=1}^N \lla F_{\alpha ij} r_{\alpha ij} \eta_{\alpha ij} \rra ,
\end{align}
an expression, which has already been derived in Ref.~\onlinecite{das:19}. Since only bulk properties are contained in this equation, it presents the equation of state for a rotational overdamped dynamics.

\section{Stress Tensor of Periodic System}

\subsection{Global stress tensor} \label{sec:stress_tensor_global_pbc}

The stress tensor of a periodic system can be derived in a similar manner. Multiplication of the equations of motion \eqref{eq:eqm_r} by $r_{\alpha i}$, averaging, and summation over all particles yields
\begin{equation} \label{eq:virial_tot}
 m \sum_{i=1}^N \lla \dot{r}^2_i \rra  - \gamma N D_t + \gamma \sum_{i=1}^N \lla v_i^{(a)} r_i  \rra + 
 \sum_{i=1}^N \lla F_i  r_i \rra   = 0 ,
\end{equation} 
with the identities $ d(\dot{\bm r}_i \cdot \bm r_i)/dt = \ddot{\bm r}_i \cdot \bm r_i + \dot{\bm r}_i^2$ and $d \bm r^2_i/dt = 2 \bm r_i \cdot \dot{\bm r}_i$.~\cite{wink:15} Here, $D_t$ is the translational diffusion coefficient, which follows from the particles' mean-square displacements.~\cite{wink:15,das:19} For pairwise forces, the force on particle $i$ is given by
\begin{equation} \label{eq:def_force}
   \bm F_i  =  \psum_{j=1}^N \sum_{\bm n} \bm F_{ij}(\bm r_i - \bm r_j - \bm R_{\bm n}) .
\end{equation}
In periodic systems, interactions between the $N$ particles may occur and, in addition, interactions with periodic images located at $\bm r_j + \bm R_{\bm n}$. \cite{wink:92.1,wink:15} For a square (cubic) periodic box of side length $L$, the lattice vector $\bm R_{\bm n}$ is given by
$\bm R_{\bm n} = L \bm n$, with $n_{\alpha} \in \mathbb{Z}$. In general, this involves infinitely many interactions.\cite{wink:02.1} Here, we assume short-range interactions only, which limits the values of $\bm n$ for any particle. \cite{wink:92.1,wink:15} With the abbreviation $\bm F_{ij}(\bm r_i - \bm r_j - \bm R_{\bm n})= \bm F_{ij}^{\bm n}$, the force term in Eq.~\eqref{eq:virial_tot} becomes
\begin{equation} \label{eq:def_force_tot}
   \sum_{i=1}^N \lla \bm F_i \cdot \bm r_i \rra  = \frac{1}{2}  \sum_{i=1}^N \psum_{j=1}^N \sum_{\bm n}\lla \bm F_{ij}^{\bm n} \cdot (\bm r_i - \bm r_j)\rra .
\end{equation}
Note that the positions $\bm r_i$, $\bm r_j$ refer to infinite space and not only the periodic volume, i.e., the ``primary'' box.~\cite{wink:92.1,wink:09,wink:15}

Multiplication of the equations of motion \eqref{eq:eqm_e} by $r_{\alpha i}$, averaging, and summation over all particles yields
\begin{equation} \label{eq:virial_rot_periodic_global}
 I \sum_{i=1}^N  \lla  \dot{e}_{i} \dot r_{i}  \rra  + \gamma_r \sum_{i=1}^N  \lla  e_{i} \dot r_{i}  \rra +   \sum_{i=1}^N  \lla \lambda_i  e_{i} r_{i}  \rra    =   0 , 
\end{equation}
with the identities $\langle d(\dot{\bm e}_i \cdot \bm r_i)/dt \rangle = \langle \ddot{\bm e}_i \cdot \bm r_i \rangle  + \langle \dot{\bm e}_i \cdot \dot{\bm r}_i \rangle =0$, and $\langle d({\bm e}_i \cdot \bm r_i)/dt \rangle = \langle \dot{\bm e}_i \cdot \bm r_i  \rangle  + \langle {\bm e}_i \cdot \dot{\bm r}_i \rangle   =0$.  

In the limit $I \to 0$, Eq.~\eqref{eq:virial_rot_periodic_global} reduces to \cite{das:19}
\begin{equation} \label{eq:swim-momentum}
  \sum_{i=1}^N  \lla  e_{i} \dot r_{i}  \rra -    \gamma_R \sum_{i=1}^N  \lla  e_{i} r_{i}  \rra    =   0 ,
\end{equation}
and the swim and swim-momentum stresses are equal. However, we emphasize that for $I > 0$ this no longer applies (cf. Eq.~\eqref{eq:virial_rot_periodic_global}). Hence, the swim stress is no longer characterizing the contribution of activity to the equation of state in the presence of rotational inertia.     

As in the derivation of Eq.~\eqref{eq:virial_local_tot}, we  multiply  Eq.~\eqref{eq:virial_rot_periodic_global} by $v_0\gamma/(\gamma_r \gamma_R)$, add and subtract the term  $\gamma \sum_i  \langle v_i^{(a)} r_i \rangle$, and replace the latter term in Eq.~\eqref{eq:virial_tot}, which yields the \REV{virial} equation 
\begin{widetext}
\begin{align}  \nonumber \label{eq:virial_tot_global} \nonumber
  m \sum_{i=1}^N \lla \dot{r}_i^2 \rra & \ - \gamma N D_t + \frac{1}{2} \sum_{i=1}^N \psum_{j=1}^N \sum_{n} \lla F_{ij}^{n}  ( r_i - r_j)\rra \\ & \  + \frac{\gamma}{\gamma_R} \sum_{i=1}^N  \lla  v^{(a)}_{i} \dot r_{i} \rra + \frac{\gamma I}{\gamma_r \gamma_R} \sum_{i=1}^N  \lla \dot v^{(a)}_{i}  \dot r_{i}  \rra  + \frac{\gamma }{\gamma_r \gamma_R}  \sum_{i=1}^N  \lla (\lambda_i + \gamma_r \gamma_R)  v^{(a)}_{i} r_{i}   \rra = 0 .
\end{align}
\end{widetext}
In the course of time, either a ``real'' or an image particle is in the primary simulation volume. Denoting the position (image or real) of a particle in the primary box by $\bm r_i'(t)$, the particle position itself is given by $\bm r_i(t) = \bm r_i'(t) + \bm R_n(t)$, where $\bm R_n(t)$ is the lattice vector at time $t$. Internal and external stress tensors can then be defined as
\begin{widetext}
\begin{align}  \label{eq:stress_int_pbc} 
V \sigma_{\alpha \alpha}^{(i)} = & \ - m \sum_{i=1}^N  \lla \dot{r}_{i \alpha}^2 \rra  -  \frac{1}{2} \sum_{i=1}^N \psum_{j=1}^N \sum_{n_{\alpha}} \lla F_{\alpha ij}^{n_{\alpha}}(r_{ij \alpha} - R_{n_{\alpha}})  \rra   - \frac{\gamma}{\gamma_R} \sum_{i=1}^N  \lla  v^{(a)}_{\alpha i} \dot r_{\alpha i} \rra  - \frac{\gamma I}{\gamma_r \gamma_R} \sum_{i=1}^N  \lla \dot v^{(a)}_{\alpha i}  \dot r_{\alpha i}  \rra , \\
V \sigma_{\alpha \alpha}^{(e)} = & \ \gamma N D_{t} - \frac{1}{2}  \sum_i \psum_j \sum_{n_{\alpha}} F_{\alpha ij}^{n_{\alpha}} R_{n_{\alpha}}  - \frac{\gamma }{\gamma_r \gamma_R}  \sum_{i=1}^N  \lla (\lambda_i + \gamma_r \gamma_R)  v^{(a)}_{\alpha i} R_{n_\alpha}   \rra ,
\end{align}
\end{widetext}
with the abbreviation $\bm r_{ij} = \bm r_i - \bm r_j$ and $\sigma_{\alpha \alpha}^{(i)} = \sigma_{\alpha \alpha}^{(e)}$.
Because $r_{\alpha i}'$ is limited and the system is homogeneous and isotropic, the average 
$\sum_{i=1}^N  \langle (\lambda_i + \gamma_r \gamma_R)  v^{(a)}_{\alpha i} r_{\alpha i}'  \rangle =0$. 
The first, second, and third terms of the internal stress tensor, $\sigma_{\alpha \alpha}^{(i)}$, represent the kinetic, interparticle force,  and swim momentum contributions. The last term arises additionally from rotational inertia. In the absence of rotational inertia, $I=0$, Eq.~\eqref{eq:stress_int_pbc} reduces to the expression provided in Ref.~\onlinecite{das:19}.

The external stress tensor comprises a contribution from particle transport, which is proportional to the translational diffusion coefficient of the active particles, as well as ``boundary'' terms due to particle interaction  with images. The diffusive term emerges from the present heat bath; it results from the friction term in the equation of motion.  

Noteworthy, the swim stress Eq.~\eqref{eq:swim_stress} does not appear in Eq.~\eqref{eq:stress_int_pbc}. Yet, in the case of an overdamped rotational dynamics, the swim stress becomes identical with the swim-momentum stress in Eq.~\eqref{eq:swim_mom_stress} (cf. Eq.~\eqref{eq:swim-momentum}).

\subsection{Local stress tensor}

The local expressions for the virials in  Eqs.~\eqref{eq:virial_r}, \eqref{eq:virial_e}, and \eqref{eq:virial_local_tot} apply to periodic systems as well. The only difference is the expression of the pairwise forces. In analogy to the derivation of the global stress tensor (Sec.~\ref{sec:stress_tensor_global_pbc}), we thus find the local internal stress tensor  
\begin{widetext}
\begin{align} \label{eq:stress_int_pbc_local}
 \Delta V \sigma_{\alpha \alpha}^{(i)}  = - m \sum_{i=1}^N \lla \dot{r}_{\alpha i}^2 \Lambda_{i} \rra -  \frac{1}{2} \sum_{i=1}^N \psum_{j=1}^N \sum_{n_{\alpha}} \lla F_{\alpha ij}^{n_{\alpha}}(r_{\alpha ij} - R_{n_{\alpha}}) \eta_{\alpha ij} \rra - \frac{\gamma}{\gamma_R} \sum_{i=1}^N  \lla  v^{(a)}_{\alpha i} \dot r_{\alpha i} \Lambda_{i} \rra - \frac{\gamma I}{\gamma_r \gamma_R} \sum_{i=1}^N  \lla  \dot v^{(a)}_{\alpha i} \dot r_{\alpha i} \Lambda_{i} \rra .
\end{align}
\end{widetext}
As for the global stress tensor (Eq.~\eqref{eq:stress_int_pbc}), the contribution $\sum_{i=1}^N  \langle (\lambda_i + \gamma_r \gamma_R)  v^{(a)}_{\alpha i} r_{\alpha i}' \Lambda_i  \rangle$ of the constraint forces and the swim stress vanishes for a homogeneous and isotropy system.

\section{Correlation Function}

To achieve theoretical insight into the various stress contributions, the correlation function for the propulsion direction of an ABP+TRI is required. 
Analytical calculations of the autocorrelation function in two and three dimensions---the latter for a sphere---yield the double-exponential decay \cite{loew:20,sand:20,sand:23}
\begin{align}\label{eq:c(t)} \nonumber
C_e(t) = & \ \lla \bm e_i(t) \cdot \bm e_i(0) \rra \\ = & \ \exp\left(-(d-1)D_r \tau_I \left[t/\tau_I- 1 + e^{-t/\tau_I} \right] \right) , 
\end{align}
where $\tau_I = I / \gamma_r$ is the angular-velocity relaxation time, and $D_r \tau_I = J$.

\begin{figure}[b]
    \centering
    \includegraphics[width = \columnwidth]{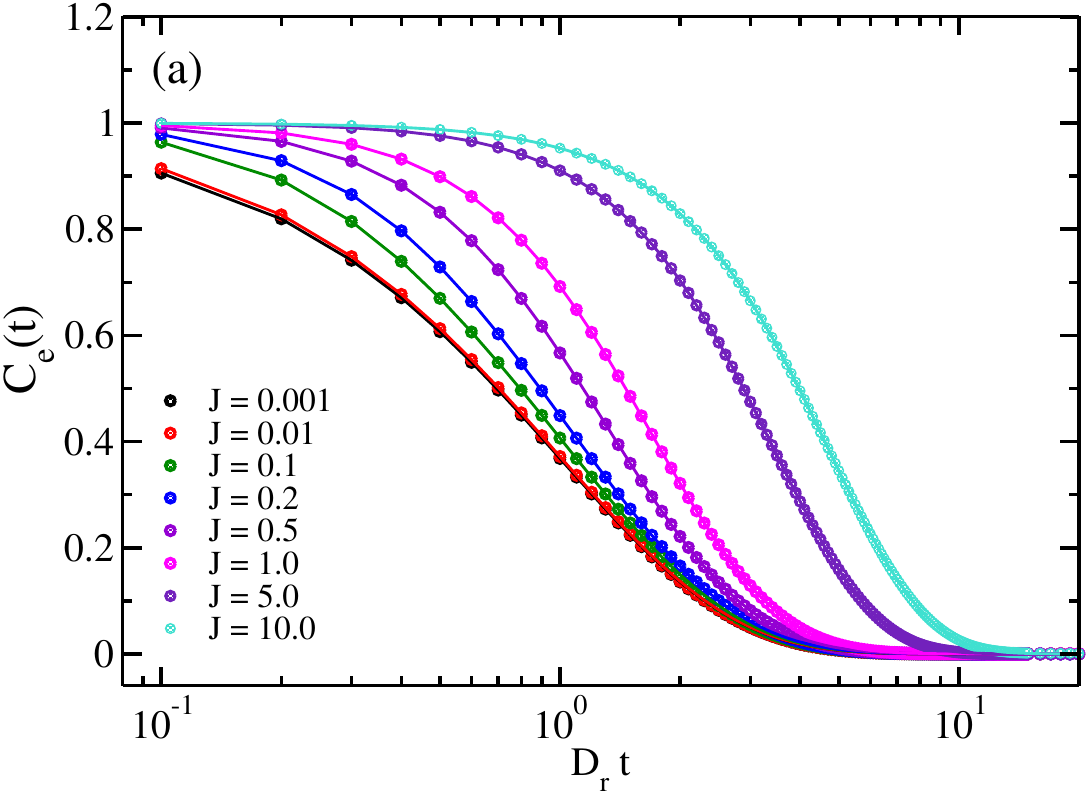}
     \includegraphics[width = \columnwidth]{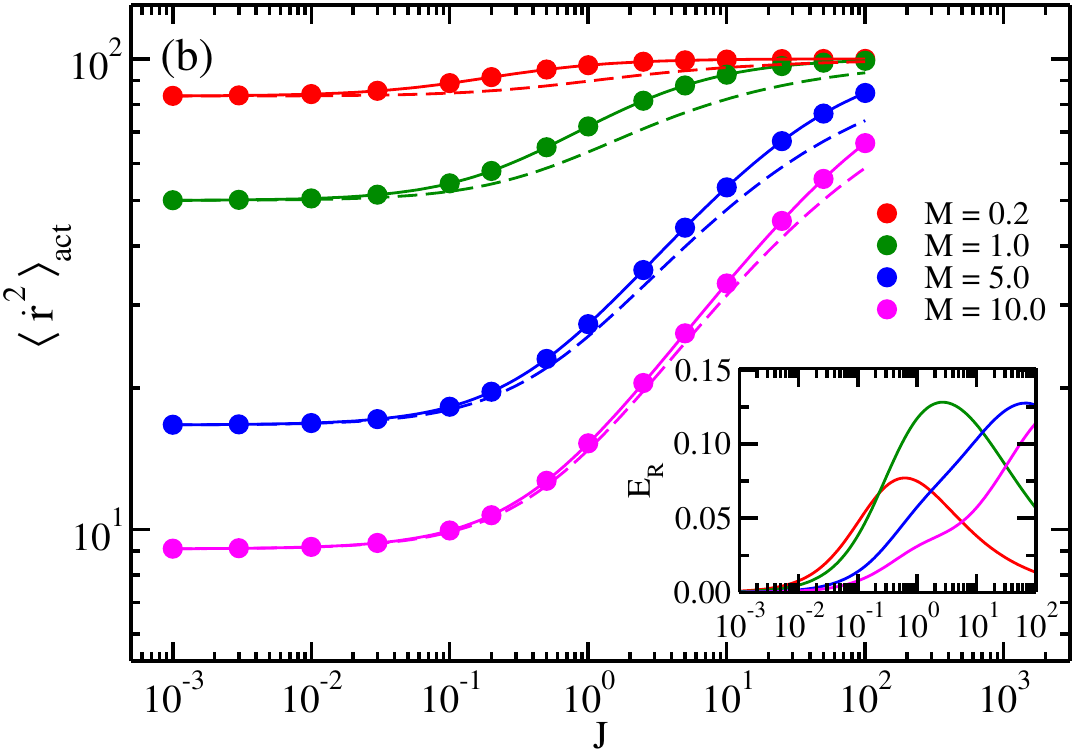}
    \caption{(a) Orientational correlation function $C_e(t)$ of ABP+TRIs as a function of  $D_r t$ for various rotational inertia $J$ (legend) in two dimension. The solid lines are the analytical expression Eq.~\eqref{eq:c(t)}~\cite{loew:20,sand:20} and the symbols are simulation results. {\cred (b) Variation of the active contribution to the mean-square velocity as a function of $J$ for various $M$. Solid and dashed lines indicate the dependence of Eq.~(15) of Ref.~\onlinecite{sand:20} and Eq.~\eqref{eq:msq_velo_act} on $J$, respectively; bullets are simulation results. The inset shows the relative error between the approximate (Eq.~\ref{eq:msq_velo_act}) and exact mean-square velocity.}}
    \label{fig:theta_correlation}
\end{figure}

Figure~\ref{fig:theta_correlation}(a) presents a comparison between the theoretical prediction and simulation results for various $J$ of a two-dimensional system ($d=2$). 
The simulation results agree with the theoretical prediction. The correlation function decreases over time and decays exponentially for $t/\tau_I \gg 1$. Importantly, the curves shift toward longer times as $J$ increases. This indicates that active particles with larger $J$ retain their orientation for a longer time, and thus, exhibit enhanced directional persistence.~\cite{sand:20}

In the limit $I \to 0$, the correlation function decays exponentially, and for $J \gg 1$ and $t/\tau_I \ll 1$ initially as a Gaussian.   
To characterize the dependence of $C_e$ on the relaxation time $\tau_I$,  we determine a characteristic relaxation time $\tau_e$ by the relation $C_e(\tau_e) = 1/e$. This yields  the well-known relation $D_r \tau_e = 1$ in the limit $J \ll 1$. \cite{wink:15} In the limit $J \gg 1$, we find the $J$-dependent relaxation time $D_r \tau_e \approx \sqrt{2J}$ in two dimensions. To bridge the gap between small and large $J$ values, we introduce the interpolation 
\begin{equation} \label{eq:tau_e}
   D_r \tau_e=  \sqrt{1+2J} .
\end{equation}

Rotational inertia implies a slower relaxation of the propulsion direction, thereby affecting the persistence length of the translational motion. Introducing a persistence length via $l_p=v_0 \tau_e$, we find for $J \gg 1$
\begin{equation} \label{eq:persistence_length}
    l_p/\sigma = Pe \sqrt{2 J} ,
\end{equation}
i.e., the persistence length increases with the square root of the rotational moment of inertia. 

Translational and rotational inertia affect the velocity correlation function $C_v(t) = \langle \dot{\bm r}_i(t) \cdot \dot {\bm r}_i(0) \rangle$ of the active particles. Approximating the correlation function $C_e$ of the propulsion direction by an exponential with the inertia-dependent relaxation time $\tau_e$, Eq.~\eqref{eq:eqm_r}  yields
\begin{equation} \label{eq:v_coor_tot}
C_v = \frac{v_0^2}{1-(\tau_m /\tau_e)^2} \left(e^{-t/\tau_e} - \frac{\tau_m}{\tau_e} e^{- t/\tau_m}  \right) +\frac{dk_BT}{m} e^{- t/\tau_m} ,
\end{equation}
where the last term on the right-hand side is the thermal contribution. Both inertial relaxation processes contribute to the decay, with the characteristic relaxation times $\tau_m$  and $\tau_e$, respectively. The mean-square velocity, i.e., $C_v(0)$ follows from Eq.~\eqref{eq:v_coor_tot} as~\cite{das:19}
\begin{equation} \label{eq:v_msq}
    \lla \dot {\bm r}_i^2 \rra = \frac{v_0^2}{1+ \tau_m /\tau_e} +  \frac{dk_BT}{m} .  
\end{equation}
(More general expressions are provided in Refs.~\onlinecite{scho:18,sand:20}.)

\REV{A comparison of the active contribution to the approximate mean-square velocity 
\begin{equation} \label{eq:msq_velo_act}
    \lla \dot {\bm r}_i^2 \rra_{\rm act} = \frac{v_0^2}{1+ M/\sqrt{1 + 2J }},
\end{equation} with that resulting from the exact correlation function~\cite{scho:18,sand:20} is displayed in Fig.~\ref{fig:theta_correlation}(b). The simulation results agree with the exact expression, and the approximation remains close to the
exact result over the full parameter range considered. As indicated by the inset, the relative error of Eq.~\eqref{eq:msq_velo_act}
with respect to the exact expression remains below approximately $12\%$ over the full range of $J$ value. This demonstrates that the approximation for $C_e$ captures the relevant behavior while retaining analytical simplicity and allows for the derivation of closed-form expressions.  
} 

In the limit $J \to 0$, i.e., $\tau_e = 1/D_r$, $\tau_m/\tau_e = m D_r/\gamma$, and the active term in Eq.~\eqref{eq:v_msq} reduces to  $v_0^2/(1+ M)$. \cite{das:19,taka:17} Vanishing translational inertia and/or a very large rotational moment of inertia, i.e., $M/\sqrt{2J} \ll 1$, imply the same limit for the active part of the mean-square velocity, namely  $v_0^2/(1+ \tau_m /\tau_e) \to v_0^2$.


\begin{figure*}[t]
    \centering
       \includegraphics[width = 0.32\textwidth]{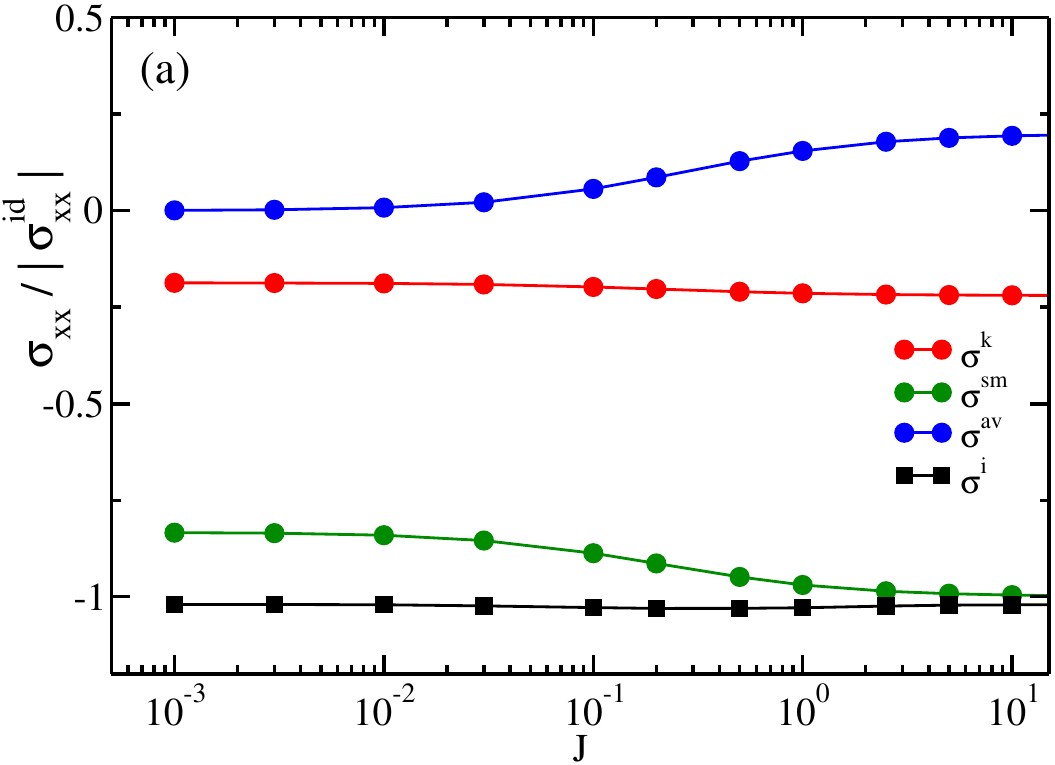}
       \includegraphics[width = 0.325\textwidth]{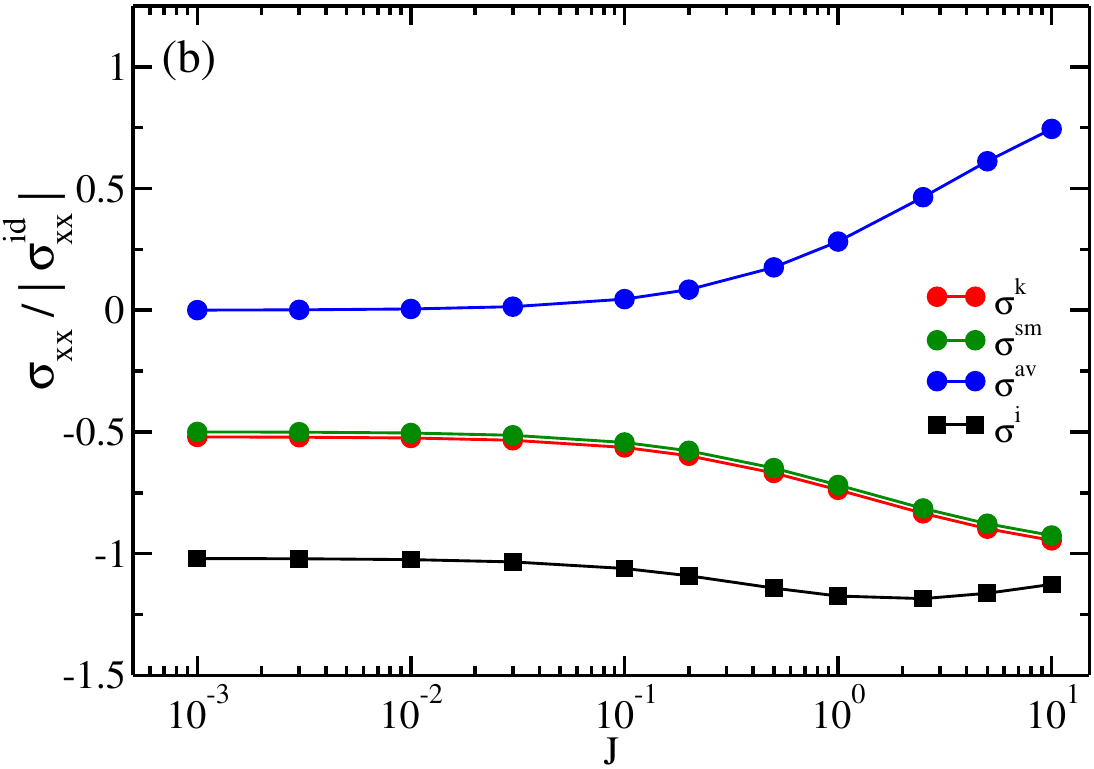}
       \includegraphics[width = 0.32\textwidth]{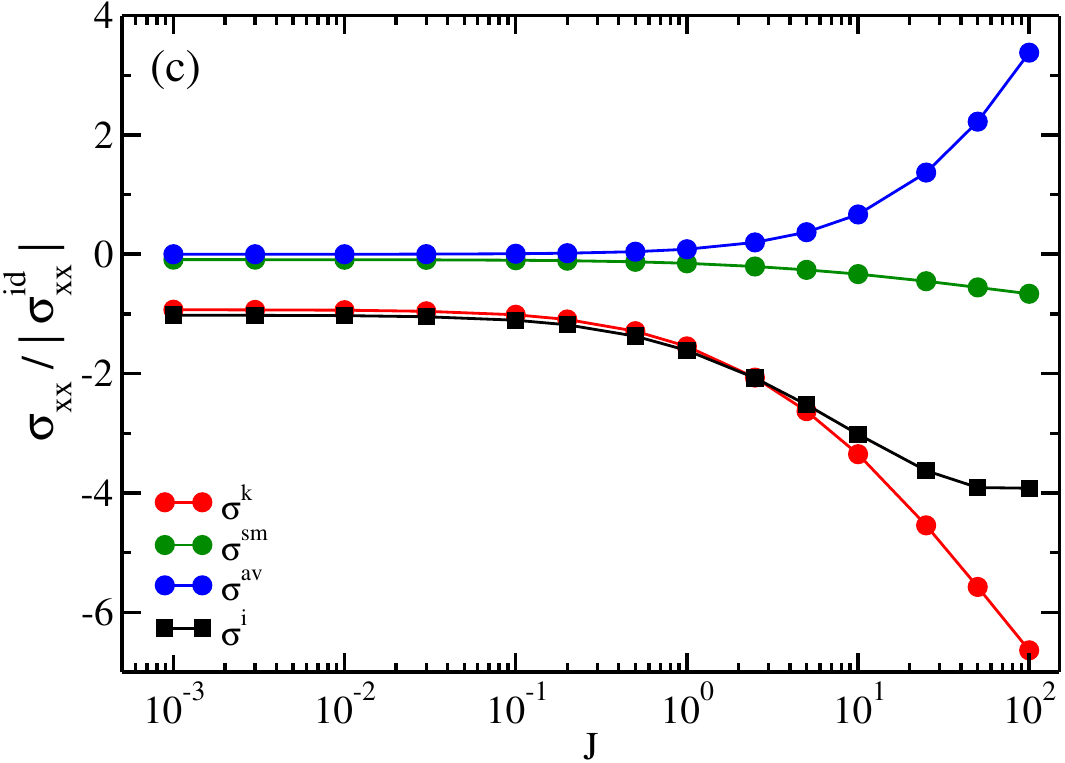}
    \caption{Virial components $\sigma_{\alpha \alpha}^{(k)}$ (red), $\sigma_{\alpha \alpha}^{(sm)}$ (green), and $\sigma_{\alpha \alpha}^{(av)}$ (blue) of the internal stress $\sigma_{\alpha\alpha}^{(i)}$ (Eq.~\eqref{eq:stress_int_pbc}) (black squares) in a periodic system for an ideal gas of active Brownian particles with translational and rotational inertia (ABP+TRIs) as a function of the reduced moment of inertia $J$ and the reduced masses (a) $M = 0.2$, (b) $M = 1.0$, and (c) $M = 10.0$ at $Pe = 10$. $\sigma_{xx}^{(id)}$ is the swim stress of an ideal ABP gas (Eq.~\eqref{eq:id_sm}). }
    \label{fig:global_comp_pbc}
\end{figure*}

The velocity correlation function \eqref{eq:v_coor_tot} yields the diffusion coefficient 
\begin{equation} \label{eq:diff_coeff_tau}
 D_t = D_0 + \frac{v_0^2 \tau_e}{2} .    
\end{equation}
It is independent of translational inertia, but depends on the rotational moment of inertia; the latter is in contrast to the conclusion in Ref.~\onlinecite{taka:17}. More precise values of the diffusion coefficient valid for all $I$ are provided in Refs.~\onlinecite{sand:20,scho:18}. 
Substituting the interpolation equation Eq.~\eqref{eq:tau_e} yields the limiting values for the diffusion coefficient
\begin{equation}
    D_t = D_0+  \left\{
    \begin{array}{cc}
         \displaystyle  \frac{v_0^2}{2 D_r} [1+J] , & J \ll 1 ,  \\[10pt]
         \displaystyle \frac{v_0^2}{2 D_r} \sqrt{2J} ,  &  J \gg 1 .
    \end{array}
    \right.
\end{equation}
The limit $J \ll 1$ is identical with the result presented in Ref.~\onlinecite{scho:18} in two dimensions. The limit $J \gg 1$ agrees with the 
asymptotic value~\cite{olve:10} of the exact expression of Refs.~\onlinecite{scho:18,sand:20,sand:23}
\begin{equation}
    D_t = D_0 + \frac{v_0^2}{2D_r} \sqrt{\frac{\pi J}{2}}
\end{equation}
within approximately $12\%$. Importantly, for $J \gg 1$, $D_t$ is independent of $M$ and increases with the square root of the moment of inertia.

\section{Ideal ABP+TRI Gas: Stress in  Periodic system}

In the following, we consider an ideal gas of ABP+TRIs in two dimensions, for which $\bm F_{ij} =0$.  This system is homogeneous and isotropic; hence, the global and local internal stress tensors are identical. Thus, we focus on the global stress tensor in this section.  

The virial expression for the global internal stress tensor of a periodic system (Eq.~\eqref{eq:stress_int_pbc}) comprises kinetic $\sigma_{\alpha \alpha}^{(k)}$ (Eq.~\eqref{eq:kin_stress}), swim momentum $\sigma_{\alpha \alpha}^{(sm)}$ (Eq.~\eqref{eq:swim_mom_stress}), and angular velocity $\sigma_{\alpha \alpha}^{(av)}$ (Eq.~\eqref{eq:ang_velo_stress}) contributions. 

Applying the approximation of an exponentially decaying propulsion-direction correlation function with the relaxation time $\tau_e$ (Eq.~\eqref{eq:tau_e}), we obtain the approximate expressions for the kinetic and swim-momentum stresses:
\begin{align} \label{eq:kin_stress_approx}
 \sigma_{\alpha \alpha}^{(k)} = & \ \sigma_{\alpha \alpha}^{(id)} \left(\frac{d \gamma_R k_BT }{\gamma v_0^2} +  \frac{M}{1+M/\sqrt{1+2J}}  \right) , \\ \label{eq:sm_stress_approx}
  \sigma_{\alpha \alpha}^{(sm)} = & \ \sigma_{\alpha \alpha}^{(id)} \frac{1}{1+M/\sqrt{1+2J}}  ,
\end{align}
with the swim stress of an ideal ABP gas \cite{taka:14,yang:14,das:19}
\begin{equation} \label{eq:id_sm}
    \sigma_{\alpha \alpha}^{(id)} = - \frac{\gamma N}{d (d-1) D_r V} v_0^2 .
\end{equation}
The derivation of an equivalent expression for the angular-velocity stress is less straightforward, as it requires the correlation between $\bm e_i(t)$ and $\dot {\bm e}_i(t)$. In the limit $J \to 0$, Eqs.~\eqref{eq:kin_stress_approx} and \eqref{eq:sm_stress_approx} agree with the prediction for the overdamped propulsion-direction dynamics.\cite{taka:17,das:19}  Moreover, the equations agree with the respective simulation results to within $15\%$.  

\begin{figure}[b]
    \centering
    \includegraphics[width = \columnwidth]{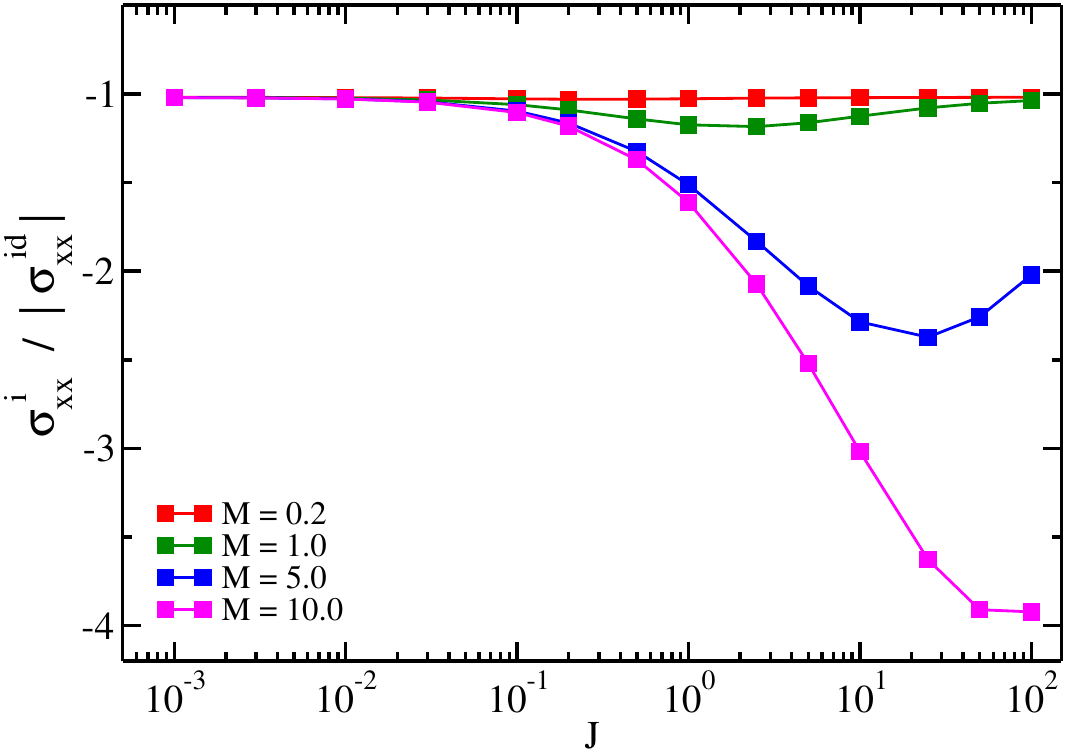}
    \caption{Internal stress tensor, $\sigma_{\alpha \alpha}^{(i)}$ (Eq.~\eqref{eq:stress_int_pbc}), for a periodic system of an ABP+TRI gas as a function of the reduced moment of inertia $J$ and the reduced masses $M = 0.2$ (red), $1.0$ (green), $5.0$ (blue), and $10.0$ (magenta) at  $Pe = 10$.}
    \label{fig:virial_int_pbc}
\end{figure}

Figure~\ref{fig:global_comp_pbc} presents simulation results for the various stress components as well as the total internal stress as a function of the reduced moment of inertia $J$ and several values of the reduced mass $M$. The moment-of-inertia-dependent stress $\sigma_{\alpha \alpha}^{(av)}$  is inherently zero in the limit $J \to 0$, but grows with increasing moment of inertia at a given $M$ and yields a positive contribution to the total stress. As a consequence, for  $J \ll 1$, the total internal stress is given by the sum of the kinetic and swim momentum stress, $\sigma_{\alpha \alpha}^{(i)} = \sigma_{\alpha \alpha}^{(k)} + \sigma_{\alpha \alpha}^{(sm)}$, and the total stress is independent of $M$ as demonstrated in Fig.~\ref{fig:virial_int_pbc}. This is the limit discussed in Refs.~\onlinecite{taka:17,sand:20,sand:23}.

The kinetic stress, $\sigma_{\alpha \alpha}^{(k)}$, is  nearly constant for $M<0.2$, but decreases with increasing $J$ for larger $M>1$, and approaches a limiting value for $J \to \infty$. The analytical expression of Eq.~\eqref{eq:kin_stress_approx} agrees well with the simulation results. This equation predicts the limiting values $M/(1+M)$ and $M$ for $J \ll 1$ and $J \gg 1$, respectively, for the $M$-dependent term. Importantly, the kinetic stress depends on $M$ as long as $M/\sqrt{1+2J} \ll 1$, however, it becomes $M$ independent for $M \gg \sqrt{1+2J}$, where $\sigma_{\alpha \alpha}^{(k)} \sim \sqrt{1+2J}$. This is apparent in Figs.~\ref{fig:global_comp_pbc}(b) and (c).  The first term on the right-hand side in Eq.~\eqref{eq:kin_stress_approx} plays a minor role, because of the quadratic dependence on the propulsion velocity $v_0$ of the active term.

The swim-momentum stress, $\sigma_{\alpha \alpha}^{(sm)}$, whose dependence on $M$ and $J$ is well described
by the analytical approximation of Eq.~\eqref{eq:sm_stress_approx}, shows a qualitatively similar behavior. It decreases monotonically with increasing $J$ and approaches the ideal active gas limit $\sigma_{\alpha \alpha}^{(id)}$ for $J \to \infty$ at any fixed $M$. The close agreement between $\sigma_{\alpha \alpha}^{(k)}$ and $\sigma_{\alpha \alpha}^{(sm)}$ for $M=1$ reflects the very similar dependence on $J$ of the two stresses as expressed in Eqs.~\eqref{eq:kin_stress_approx} and \eqref{eq:sm_stress_approx}. However, with increasing $J \gtrsim 1$, the stresses $\sigma_{\alpha \alpha}^{(k)}$ and $\sigma_{\alpha \alpha}^{(av)}$ start to compensate each other and their sum becomes zero in the limit $J \to \infty$. The total stress is then given by the swim momentum stress $\sigma_{\alpha \alpha}^{(sm)}$. 

Figure~\ref{fig:virial_int_pbc} illustrates the dependence of the internal stress tensor $\sigma_{\alpha \alpha}^{(i)}$ on $M$. The internal stress decreases above a certain value as  $J$ increases, reaches a minimum, grows again, and finally approaches a limiting value. This dependence becomes more pronounced with increasing $M$. Due to the (partial) cancellation of $\sigma_{\alpha \alpha}^{(k)}$ and  $\sigma_{\alpha \alpha}^{(av)}$, $\sigma_{\alpha \alpha}^{(i)}$ does not grow with $\sqrt{J}$ as predicted in Ref.~\onlinecite{sand:20}. Rather, the limiting value for $J \to \infty$ is given by $\sigma_{\alpha \alpha}^{(i)} = \sigma_{\alpha \alpha}^{(sm)} = \sigma_{\alpha \alpha}^{(id)}$.

\begin{figure}[t]
        \includegraphics[width=\columnwidth]{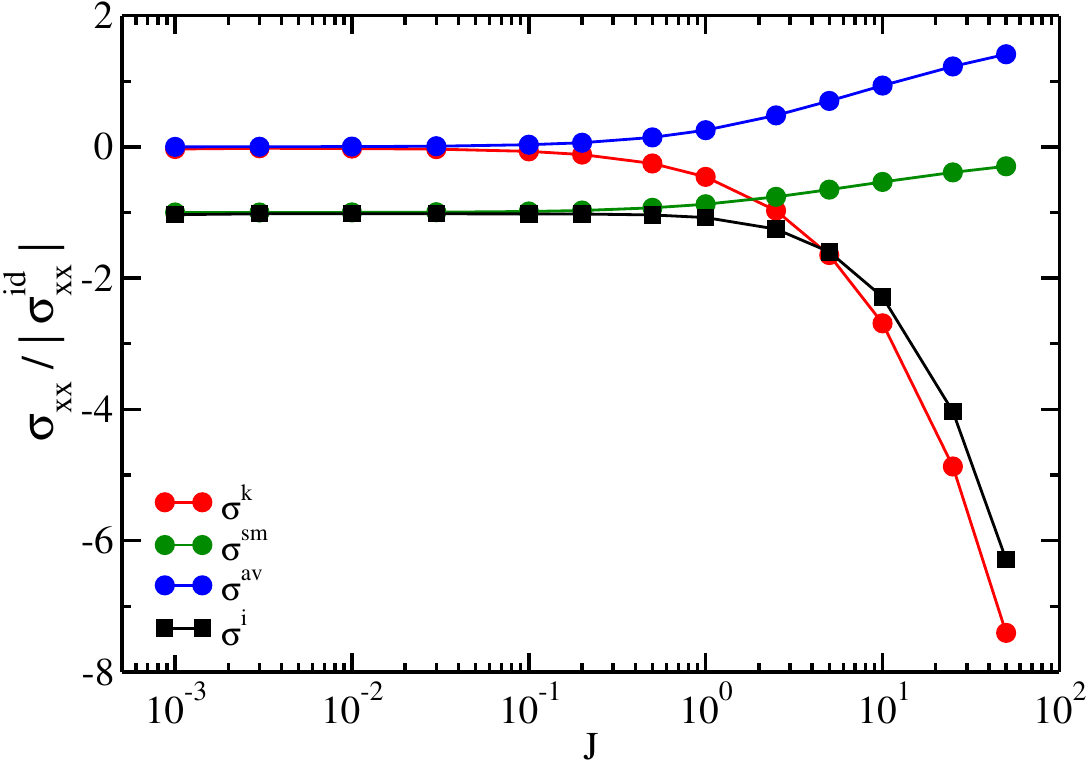}
    \caption{Internal stress $\sigma_{\alpha\alpha}^{(i)}$ (Eq.~\eqref{eq:stress_int_pbc})  (black squares) and its components  $\sigma_{\alpha \alpha}^{(k)}$ (red), $\sigma_{\alpha \alpha}^{(sm)}$ (green), and $\sigma_{\alpha \alpha}^{(av)}$ (blue)  for an ideal gas of ABP+TRIs in a periodic system as a function of the reduced moment of inertia $J$ at $Pe = 10$. The ratio of the reduced moment of inertia and the reduced mass is fixed at $J/M =2$; the $x$-axis also presents the dependence on $M$.  
     }
    \label{fig:int_stress_m_pbc}
\end{figure}

To illustrate the dependence of the internal stress and its various components on the ratio $J/M$, Fig.~\ref{fig:int_stress_m_pbc} presents results for the ratio $J/M=2$. Because of the fixed ratio, the figure also presents the dependence on the reduced mass $M$. At $J \sim M \ll 1$, the kinetic stress is very small, and the angular-velocity stress is zero. The internal stress is determined by the swim-momentum stress and equals the ideal gas stress, as expected for overdamped dynamics.~\cite{taka:14,wink:15} As $J$ and $M$ increase, the swim-momentum stress gradually vanishes and \REV{$\sigma_{\alpha \alpha}^{(k)}$ decreases.} This is well described by Eqs.~\eqref{eq:kin_stress_approx} and \eqref{eq:sm_stress_approx}. At the same time, the angular-velocity stress slowly grows, yet the total stress is essentially given by the kinetic stress. The latter exhibits the asymptotic behavior $\sigma_{\alpha \alpha}^{(k)}/\sigma_{\alpha \alpha}^{(id)} \sim \sqrt{M}$ for $M\gg1$.     

In summary, we derive the internal global and local stress tensors for an ideal ABP+TRI gas with periodic boundary conditions. The two representations yield the same stress; hence, there is an equation of state in such systems. 
Remarkably, the sum of swim and kinetic stress does not provide the correct total stress. Thus, a stress equation including the swim stress does not provide a suitable equation of state in presence of rotational inertia.  This is expressed by Eq.~\eqref{eq:virial_rot_periodic_global}, which contains, aside from the swim and swim-momentum stress, contributions including the moment of inertia. The latter points toward an augmented expression for the (local) momentum conservation compared to systems in the absence of a moment of inertia.~\cite{fily:18} Hence, the swim stress and swim-momentum stress are no longer equivalent, as in the case $J=0$. \cite{fily:18,das:19} This is in line with the results presented in Ref.~\onlinecite{fily:18} for systems, which violate local momentum conservation.       

\begin{figure*}[t]
      \centering   \includegraphics[width=0.32\textwidth]{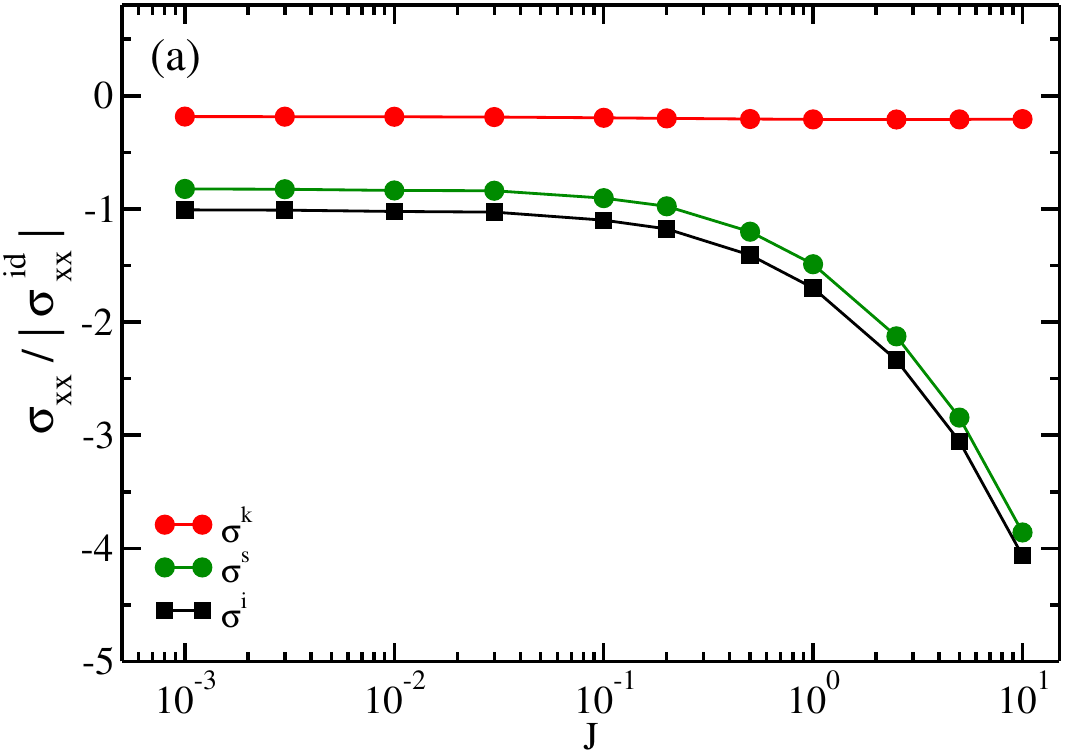}
    \includegraphics[width=0.32\textwidth]{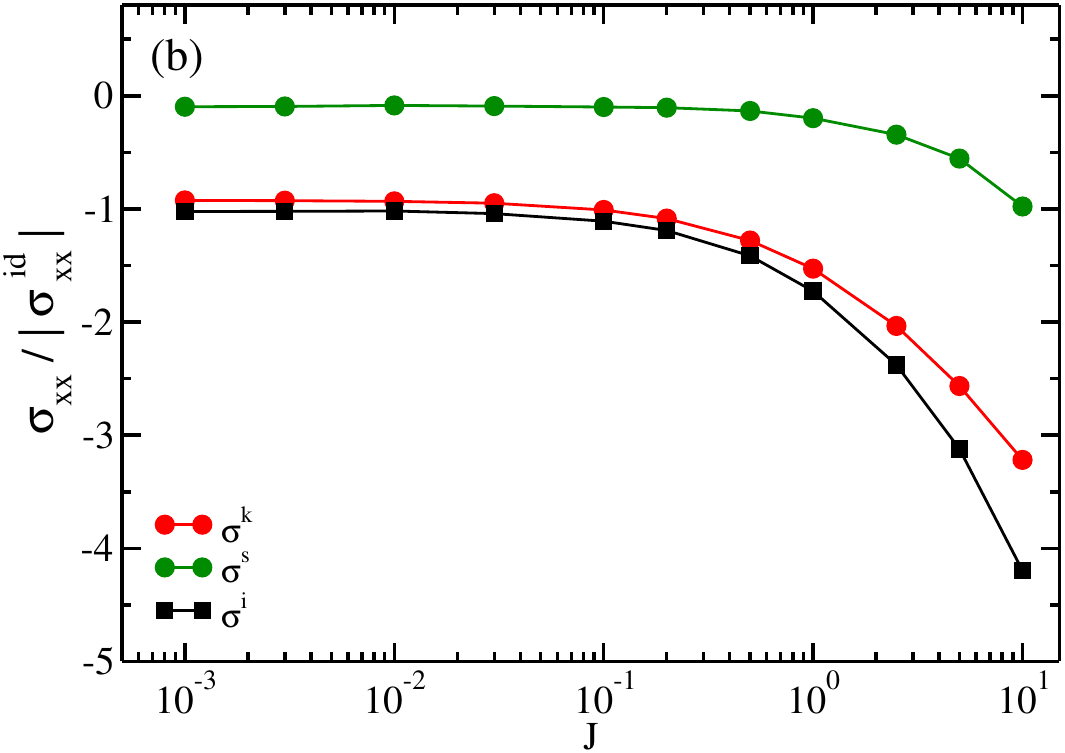}
        \includegraphics[width=0.315\textwidth]{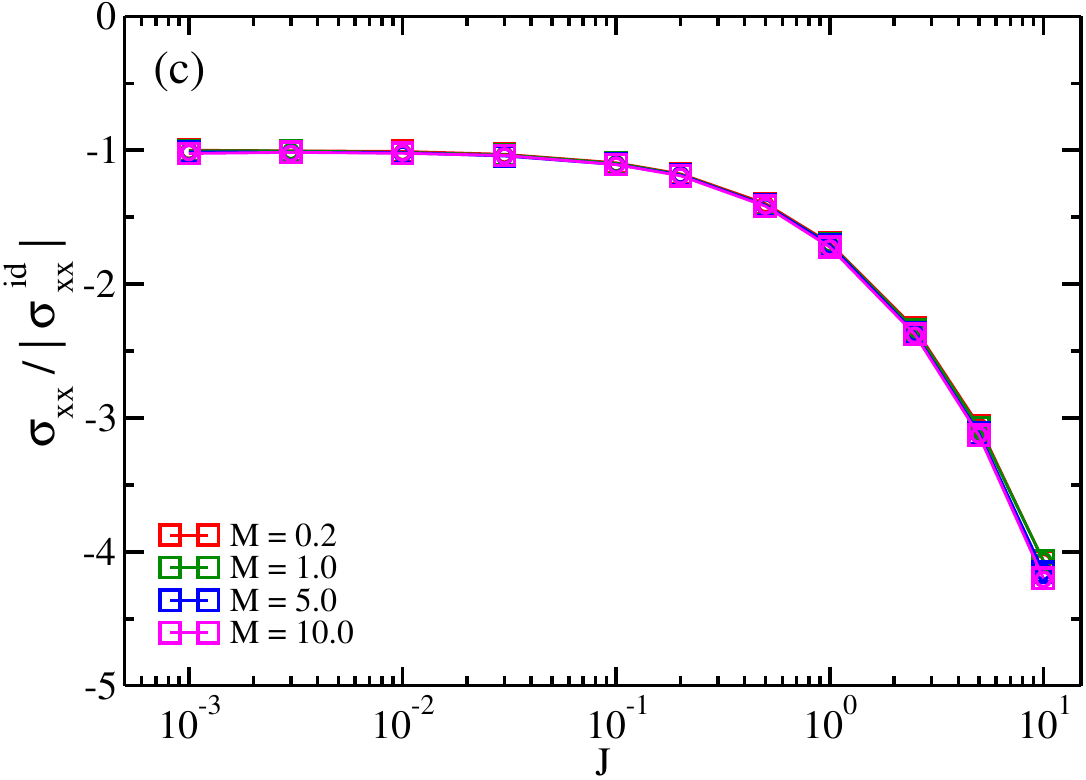}
    \caption{The virial components $\sigma_{\alpha \alpha}^{(k)}$ (red) and $\sigma_{\alpha \alpha}^{(s)}$ (green) of the global internal stress $\sigma_{\alpha\alpha}^{(i)}$ (Eq.~\eqref{eq:stress_int_conf}) (black squares) in a confined system of ABP+TRIs as a function of the reduced moment of inertia $J$ and the reduced masses (a) $M = 0.2$ and (b) $M = 10.0$ at $Pe = 10$. (c) Global internal (squares) and external (circles) stresses as a function of the reduced moment of inertia and various masses (legend). }
    \label{fig:glob_strs_conf}
\end{figure*}

\section{Ideal ABP+TRI Gas: Stress in Confined System} \label{sec:confined_sim}

In this section, we consider a two-dimensional ideal ABP+TRI gas, which is confined by ``walls''  along the $x$-axis located at positions $x_w=\pm L/2$, while periodic boundary conditions are applied along the $y$-axis. This setup enables us to directly probe the stress exerted at the walls (Eq.~\eqref{eq:stress_ext_conf}) by the gas particles and compare it with the internal global (Eq.~\eqref{eq:stress_int_conf}) and local (Eq.~\eqref{eq:local_virial_conf}) stresses in the system.

\subsection{Global stress} \label{sec:confined_global}

The stress along a confined spatial direction is given by the Eqs.~\eqref{eq:stress_ext_conf} and \eqref{eq:stress_int_conf}. Equivalently, the latter can be expressed by Eq.~\eqref{eq:virial_conf_int}. Within the approximation of an exponentially decaying propulsion-direction correlation, the swim stress Eq.~\eqref{eq:swim_stress} becomes
\begin{equation}
    \sigma_{\alpha \alpha}^{(s)} =\sigma_{\alpha \alpha}^{(id)} \frac{\sqrt{1+2J}}{(1+M/\sqrt{1+2J})} .
\end{equation}
The sum of the kinetic stress (Eq.~\eqref{eq:kin_stress_approx}) and the swim stress,
\begin{equation}
    \sigma_{\alpha \alpha}^{(k)}  + \sigma_{\alpha \alpha}^{(s)} = \sigma_{\alpha \alpha}^{(id)} \left(\frac{d \gamma_R k_BT }{\gamma v_0^2} +  \sqrt{1+2J}  \right) ,
\end{equation}
is independent of the ABP mass, consistent with the prediction in Refs.~\onlinecite{taka:17,sand:20}. Moreover, the sum of the stresses decreases as $\sqrt{J}$ for $J \gg 1$, in agreement with the finding in Ref.~\onlinecite{sand:20}. 

Figure~\ref{fig:glob_strs_conf} displays the dependence of the kinetic stress and swim stress on the moment of inertia for various masses. For a small mass $M \lesssim 0.2$, the internal stress is dominated by the swim stress, whereas for large $M>1$, the kinetic stress dominates. As shown in Fig.~\ref{fig:glob_strs_conf}(c), the sum of the two terms is independent of $M$.  Moreover, this figure illustrates the quantitative agreement between the external stress, Eq.~\eqref{eq:stress_ext_conf}, and the internal stress, Eq.~\eqref{eq:stress_int_conf}, which must be valid. The contribution of the surface term in Eq.~\eqref{eq:stress_int_conf} is very small and thus is negligible.

In general, the dependence of the global stress on the moment of inertia differs qualitatively and quantitatively from the stress in a periodic system (cf. Figs.~\ref{fig:virial_int_pbc} and \ref{fig:glob_strs_conf}(c)). In particular, the internal stress in a periodic system saturates in the limit $J/M\gg 1$, whereas in a confined system the internal global stress decreases as $\sqrt{J}$. 
Only over a certain range of moments of inertia, both expressions agree, but not for $J/M \gg 1$. This questions the existence of an equation of state in confined systems at large moments of inertia. 
As we will show in Sec.~\ref{sec:confined_local}, the reason is the emergent strong polarization of the propulsion direction in the vicinity of the walls.    

\begin{figure}[t]
    \centering
    \includegraphics[width = 0.98\columnwidth]{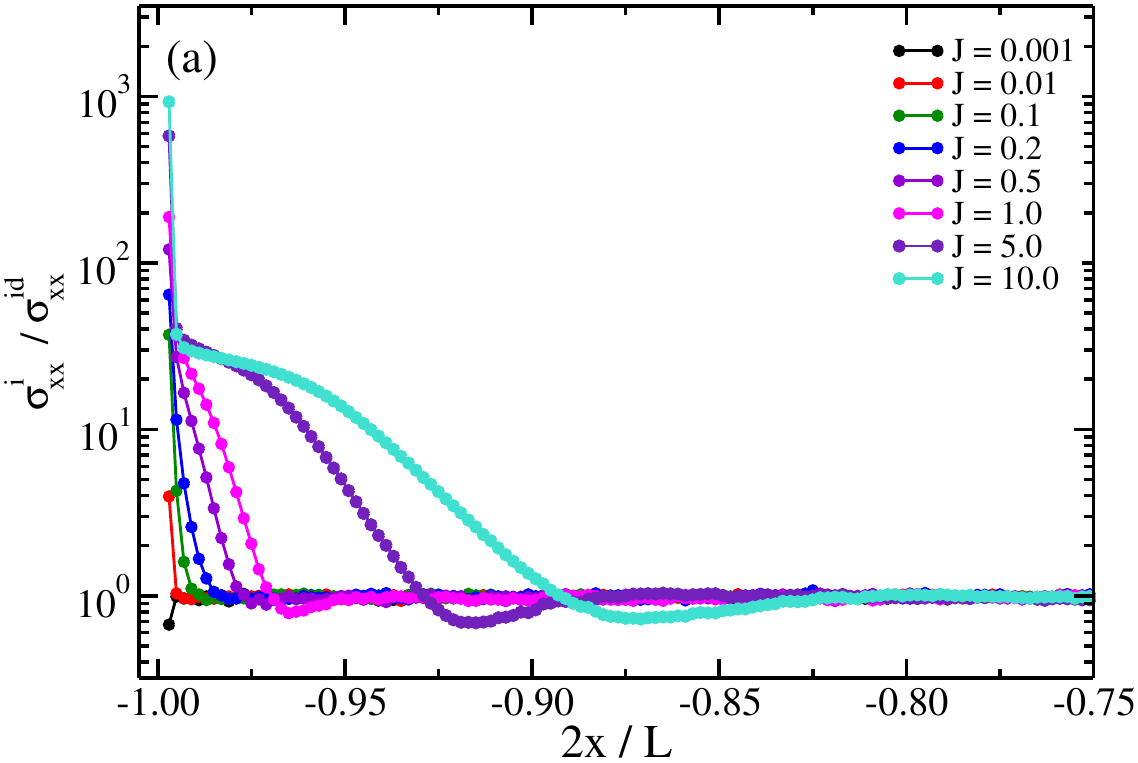}
     \includegraphics[width = 0.98\columnwidth]{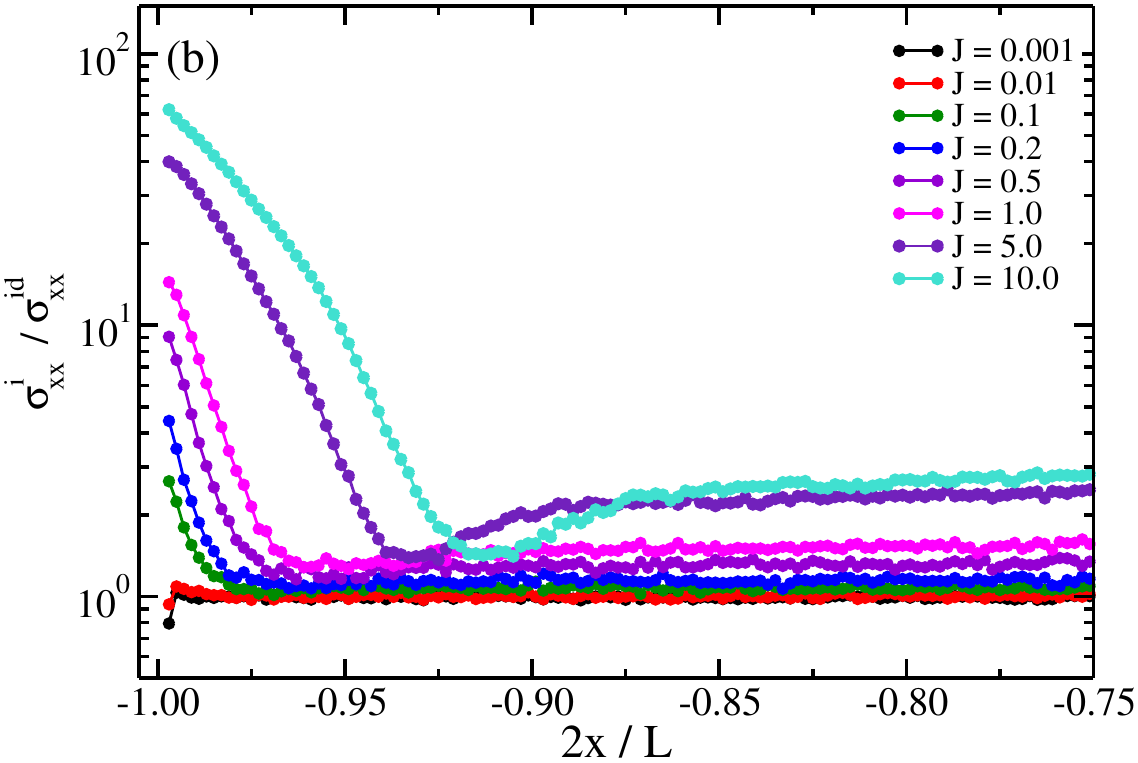}
    \caption{Internal local stress tensor (Eq.~\eqref{eq:local_virial_conf}) of a confined ABP+TRI gas as a function the scaled position $x/(L/2)$ normal to the walls for various moments of inertia $J$ (legend). In (a), the ABP mass is $M = 0.2$, and in (b), $ M = 10$. The P\'eclet number is $Pe=10$ and the wall separation $L=10^3\sigma$. }
    \label{fig:tot_strs}
\end{figure}

\begin{figure}[t]
    \centering
    \includegraphics[width = 0.98\columnwidth]{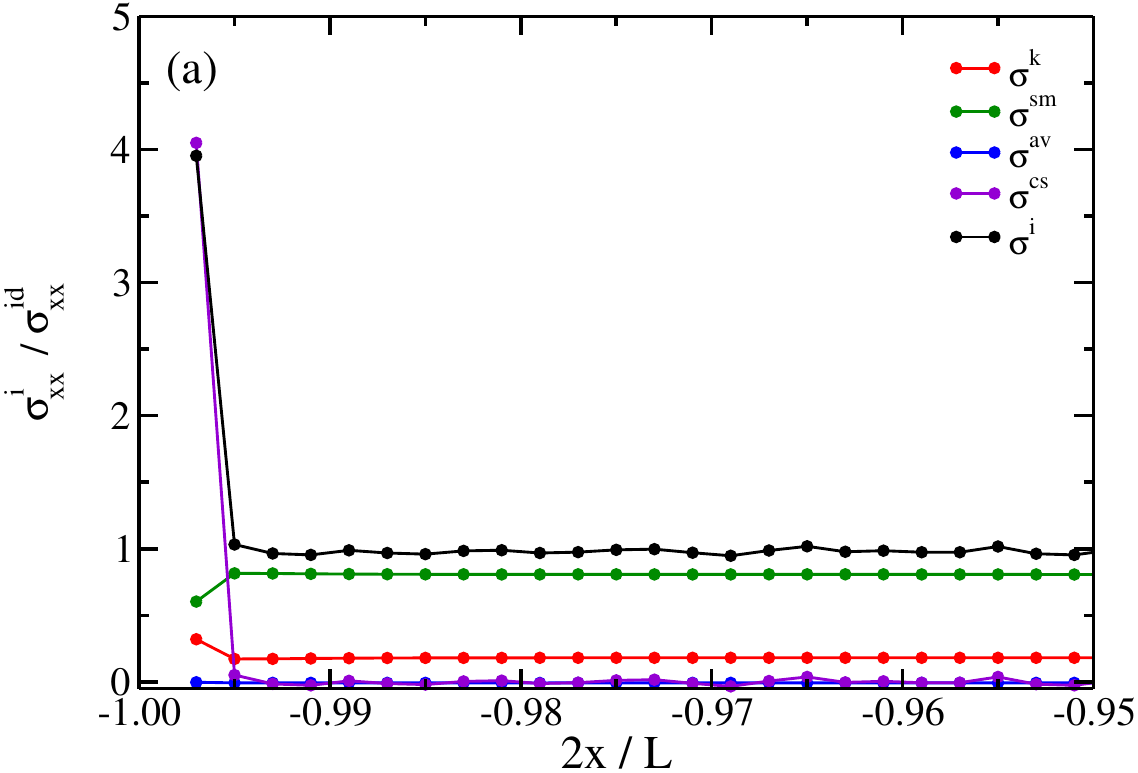}
     \includegraphics[width = 0.98\columnwidth]{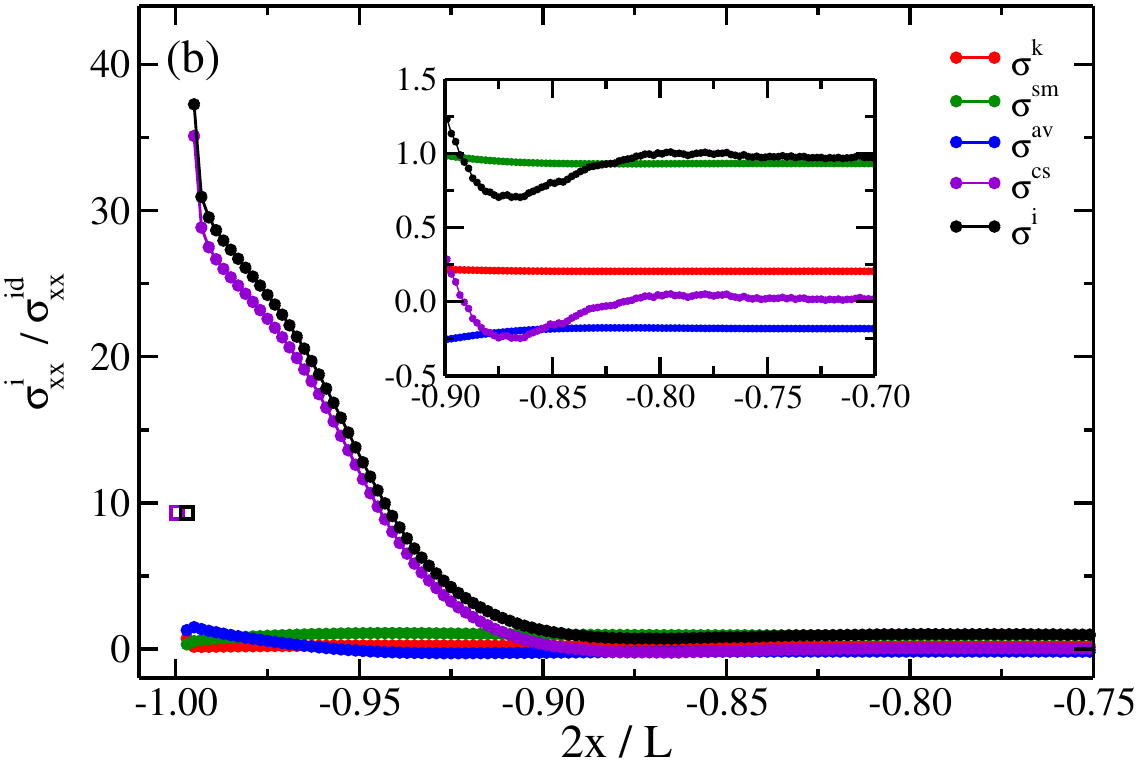}
    \caption{Local stress tensor components (Eq.~\eqref{eq:local_virial_conf}) of a confined ABP+TRI gas as a function of the scaled position $x/(L/2)$ normal to the walls for the moments of inertia (a) $J=0.01$ and (b) $J=10$. The superscripts of the stress components in the legends refer to the local stresses corresponding to Eqs.~\eqref{eq:kin_stress}, \eqref{eq:swim_mom_stress}-\eqref{eq:const_swim_stress}. In (b), the maximum values, indicated by squares, are divided by 100. Inset: Stress components in the transition region toward the bulk. 
    The mass is $M= 0.2$, the P\'eclet number $Pe=10$, and the wall separation $L=10^3\sigma$. }
    \label{fig:stress_comp}
\end{figure}

\subsection{Local stress} \label{sec:confined_local}

The internal local stress of a confined ideal ABP+TRI gas is given in Eq.~\eqref{eq:local_virial_conf}. It comprises the local contributions from the kinetic, swim-momentum, angular-velocity, and the constraint+swim stresses. Figure~\ref{fig:tot_strs} illustrates the position dependence of the local internal stress for several moments of inertia, $J$, and masses, $M$. The stress profile is symmetric with respect to the center between the two walls and assumes a plateau value sufficiently far from a wall. Hence, only one quarter of the profile next to the wall at $x_w=-L/2$ is shown. The figure reveals strong spatial variations in stress, with values adjacent to a wall being orders of magnitude larger than the bulk and global stress values. However, the bulk values for the various $J$ values agree with the stresses in the periodic system in Fig.~\ref{fig:virial_int_pbc}, taking into account the slightly reduced density in the bulk part of the confined system. Hence, the system is isotropic in the bulk regime, but is anisotropic within a moment-of-inertia-dependent surface layer. 

Figure~\ref{fig:stress_comp} displays the various stress components for the moments of inertia $J=0.01$ and $J=10$, and the mass $M=0.2$. For $J=0.01$ (Fig.~\ref{fig:stress_comp}(a)), the various components are constant, except within a surface layer extending over the wall-potential range. This layer broadens substantially with increasing $J$ and $M$ as evident from Figs.~\ref{fig:tot_strs} and \ref{fig:stress_comp}(b). For $J=0.01$ (Fig.~\ref{fig:stress_comp}(a)), the components of the stress tensor $\sigma_{xx}^{(av)}$ and $\sigma_{xx}^{(cs)}$ are equal to zero, and the internal stress is determined by the kinetic, $\sigma_{xx}^{(k)}$, and swim momentum, $\sigma_{xx}^{(sm)}$, stress, just as for the periodic system (Fig.~\ref{fig:global_comp_pbc}(a)). In the case of $J=10$ (Fig.~\ref{fig:stress_comp}(b)), $\sigma_{xx}^{(cs)}$ is still zero, but $\sigma_{xx}^{(av)}$ becomes nonzero in the bulk, as for the periodic system (cf. inset of Fig.~\ref{fig:stress_comp}(b)). The internal stress is dominated by $\sigma_{xx}^{(sm)}$, since the components $\sigma_{xx}^{(k)}$ and $\sigma_{xx}^{(av)}$ almost cancel each other out.

When the ABP+TRIs approach a wall, the magnitude of the stress component $\sigma_{xx}^{(cs)}$ increases significantly and dominates the local internal stress. This is related to an emergent polarization, i.e., the preferred orientation of the propulsion direction normal to the walls (see Sec.~\ref{sec:polarization}). \cite{yan:15,ezhi:15,elge:15,fily:18,ausc:21,carr:24} In the bulk, $\sigma_{xx}^{(cs)}$ is zero, because the average over the propulsion direction is zero and no polarization is present. The (very) large values of $|\sigma_{xx}^{(cs)}|$ at a wall are related to two aspects: the system-size dependence of $|\sigma_{xx}^{(cs)}|$ and the increased particle density at a wall. 

Rotational inertia can be a momentum source, which then implicates a violation of local momentum conservation.~\cite{fily:18} Hence, the presence of the swim stress---a nonlocal quantity---implies the absence of an equation of state in a confined system in the presence of a finite moment of inertia. \REV{Yet, we emphasize that the bulk stress is isotropic and, accounting for the small density  decrease, agrees with the stress in a periodic system.} The difference in the external stress, i.e., the force per area, is a consequence of the polarization in the vicinity of a wall. Moreover, there is an equation of state in the limit $J \to 0$, as is evident from Figs.~\ref{fig:tot_strs} and \ref{fig:stress_comp}(a) and Eq.~\eqref{eq:local_virial_conf}, which includes only local quantities.~\cite{fily:18,das:19}

\subsection{Local Density} \label{sec:density}

\begin{figure}[t]
    \centering
    \includegraphics[width = \columnwidth]{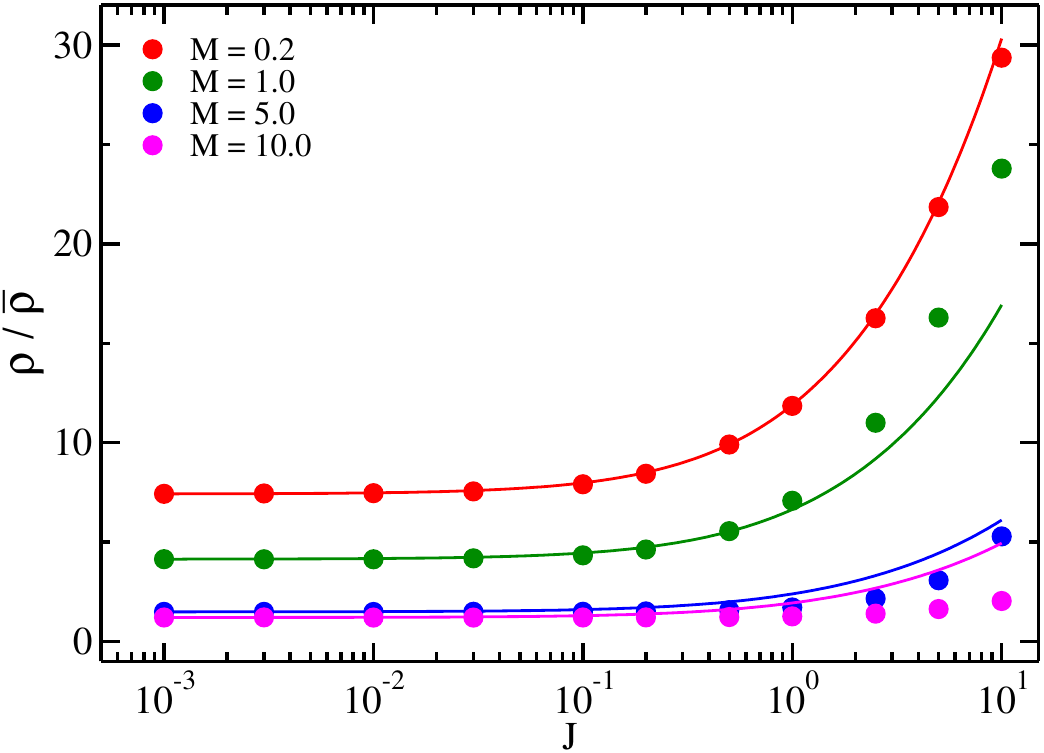}
    \caption{Particle density in the surface layer as a function of the moment of inertia and for various masses (legend). The density is normalized by the number density $\bar\rho =0.318/\sigma^2$. The solid lines present the relation $\sqrt{1 + \pi J / 2}$. The P\'eclet number is  $Pe = 10$ and the system's size is $L=10^3 \sigma$.}
    \label{fig:density}
\end{figure}

The active particles accumulate at a wall and stay there over a time interval $\tau_e \approx \sqrt{1+2J}/D_r $ as long as the active momentum of Eq.~\eqref{eq:active_moment} exceeds the translational momentum of a particle. Hence, the residence time increases with the moment of inertia, which leads to enhanced accumulation with increasing $J$. Figure \ref{fig:density} displays the density $\rho = \sum_i \Theta((L+\sigma)/2 - r_{x i})/(L \sigma)$ of particles at the surface ($\Theta (x)$ is the Heaviside step function). Already for $J=0$, the density at a wall is enhanced, \cite{elge:15,bech:16,das:19.1} to an extent that depends on the translational inertia, i.e., $M$. As anticipated, this increase is well described by $\sqrt{1+\pi J/2}$ at small $M < 1$, which is the more accurate expression for the relaxation time $\tau_e$. \cite{scho:18,sand:20} With increasing translational inertia, the density at a wall decreases for all $J$, and the increase of $\rho$ with increasing $J$ shifts to large moments of inertia. When $M \gg J$, the density becomes independent of $J$ and close to the value of a passive system. \cite{sand:23}

\begin{figure}[t]
    \centering
    \includegraphics[width = 0.98\columnwidth]{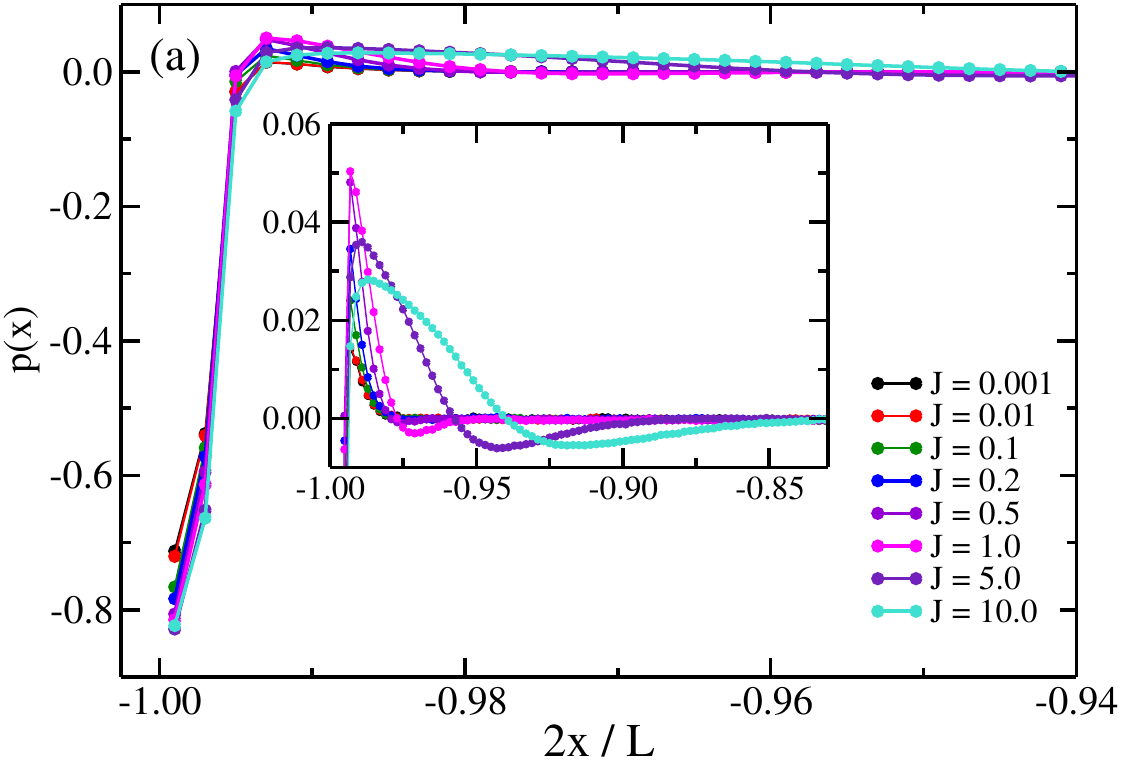}
 \includegraphics[width = 0.98\columnwidth]{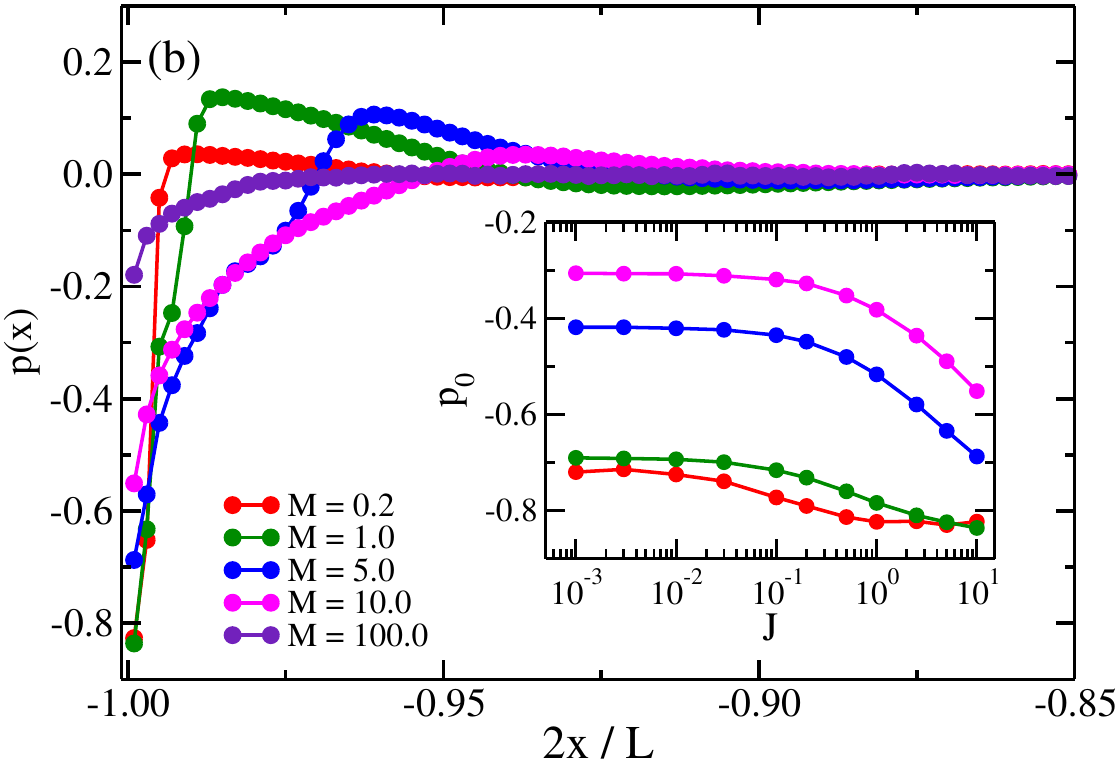}
    \caption{Local polarization $p(x)$ as a function of the particle position normal to the wall located at $x_w=-L/2$. (a) Polarization for various moments of inertia $J$ (legend) at $M=0.2$. Inset: zoomed-in view. (b) Polarization for various $M$ (legend) and $J=10$. The P\'eclet number is  $Pe = 10$ and the system's size $L=10^3 \sigma$. Inset: polarization at the surface as a function of $J$ and for various $M$. }
    \label{fig:polarization}
\end{figure}

\subsection{Local Polarization} \label{sec:polarization}

The local polarization is defined as 
\begin{equation}
    p(x) = \frac{1}{\Delta N(x)} \sum_{i=1}^N \lla e_{xi} \rra \Lambda_i(x) , 
\end{equation}
with the number of particles $\Delta N(x) = \sum_{i=1}^N  \Lambda_i(x)$ in the volume (area) $\Delta V$. Figure~\ref{fig:polarization} displays 
polarizations for various moments of inertia and masses. In general, $p(x)$ is zero in the bulk and assumes a negative value at the wall at $x_w=-L/2$. The negative polarization indicates a preferred orientation of the APB+TRI toward the wall. Further apart from a wall---the actual distance range depends on $J$ and $M$---, $p(x)$ becomes positive. This regime is dominated by particles moving away from the wall. For even large distances, the polarization is negative again before the zero bulk value is assumed. As shown in Fig.~\ref{fig:polarization}(a), the $x$-axis range of nonzero polarization broadens with increasing $J$. This is related to the increasing persistence length as $J$ increases (Eq.~\eqref{eq:persistence_length}). 

An increase in translational inertia reduces the polarization at a wall as illustrated in Fig.~\ref{fig:polarization}(b), specifically in the inset. 
Moreover, for $M/J \gg 1$, the range of a nonzero polarization is reduced. The reduction in the influence of rotational inertia on polarization and wall density is a consequence of the dominant contribution of the translational-momentum exchange with a wall compared to the contribution of the swim momentum.~\cite{loew:20} Consequently, the contribution of local stress in the proximity of walls diminished, and the stress becomes equal to that in the bulk.

\section{Summary and Conclusions}

In this article, we derive local and global internal stress tensors for systems of active Brownian particles with translational and rotational inertia (ABP+TRIs). Moreover, to evaluate the attained expressions, we have performed simulations of two-dimensional ideal ABP+TRI gases under periodic boundary conditions and confined between walls. In every case, the internal stress tensor contains a term that explicitly depends on the moment of inertia. Moreover, the swim stress does not account for the active stress in the presence of rotational inertia, in contrast to the discussions in Refs.~\onlinecite{sand:20,sand:23}. Only in the limit of vanishing rotational inertia, the swim stress correctly accounts for the active stress in periodic systems, but not for the local stress in confined systems.~\cite{das:19} Independent of the boundary condition, the swim stress vanishes locally in a homogeneous and isotropic system. For systems with periodic boundary conditions, there is an equation of state for an ideal gas of ABP+TRIs. On the contrary, walls lead to a pronounced polarization of the propulsion direction; thus, there is no equation of state in such a system.~\cite{fily:18}  However, the bulk stress (sufficiently) far from a confining wall is equal to the stress in a periodic system at the same density. Even more, in the limit of vanishing reduced moment of inertia, $J$, there is an equation of state also for confined systems.~\cite{das:19}      

Rotational inertia increases the persistence of the motion in the propulsion direction, thereby enhancing surface accumulation and polarization. Translational inertia counteracts this effect by reflection of particles back into the bulk. As a result, active systems with large $M$ exhibit reduced surface accumulation and polarization, leading to better agreement between local and global stress. Yet, a nonzero moment of inertia, in combination with a wall, acts as a local momentum source. As a consequence, there is no equation of state in such a system.~\cite{fily:18} 
 
The basic prerequisite for deriving an equation of state for dry active systems is the suitable consideration of activity. 
Extending the mechanical stress equation of passive systems, consisting of the kinetic stress and the virial of the internal forces, by the virial of the active force (swim stress) is inappropriate in general. The calculation of a local swim stress in a confined system already reveals the inappropriateness of such an approach, as this swim stress is zero.~\cite{spec:16,das:19} Our approach shows that a suitable account of activity is even more crucial for ABP+TRIs. We explicitly take into account virials emerging from the equation of motion of the propulsion direction.  By suitably including them into the virial equation of the particles' translational motion, we define the internal (local) stress tensors of Eq.~\eqref{eq:local_virial_conf} for confined, and Eqs.~\eqref{eq:stress_int_pbc} and \eqref{eq:stress_int_pbc_local} for periodic systems. These expressions contain the (local) swim-momentum stress, Eq.~\eqref{eq:swim_mom_stress} and the (local) angular-velocity stress, Eq.~\eqref{eq:ang_velo_stress}. The appearance of the constraint+swim stress, Eq.~\eqref{eq:const_swim_stress}, in the local stress of confined systems, Eq.~\eqref{eq:local_virial_conf}, reflects a possible inhomogeneity by the presence of walls, and, as mentioned before, the emergence of a local polarization. In a homogeneous and isotropic system, this term is zero. Our extension yields an equation of state for periodic systems of an ABP+TRI gas, and the local internal stress in the bulk part of a confined system agrees with the stress in the periodic system at the same density. Moreover, it yields an equation of state for (overdamped) ABPs and local stress tensors, in case of periodic boundary conditions as well as confinement, even in the presence of excluded-volume interactions.~\cite{das:19,fily:18}              


Our results highlight the interplay between translational and rotational inertia in shaping the mechanical properties of active Brownian particles and provide a more comprehensive understanding of stress, stress generation, and spatial heterogeneity in active matter systems.

\begin{acknowledgments}

SPS acknowledges funding support from the DST-SERB Grant No. CRG/2020/000661. CHT acknowledges UGC India for the fellowship and HPC facility at IISER Bhopal and Param Himalaya NSM facility.
\end{acknowledgments} 

\section*{AUTHOR DECLARATIONS}

\section*{Conflict of Interest}

The authors have no conflicts of interest to declare.


\clearpage
\newpage 

\begin{appendix}

\section{Integration of the equations for the rotational dynamics} \label{app:integ_e}

To solve equation~\eqref{eq:eqm_e} for particle $i$, the integration scheme for stochastic differential equations of Ref.~\onlinecite{gron:13} (Eqs.~(19), (20)) is applied, combined with an adaptation of the RATTLE algorithm~\cite{ande:83} to capture the constraint (the index $i$ is suppressed in the following). The integration scheme of Ref.~\onlinecite{gron:13} yields
\begin{widetext}
\begin{align} \label{eq:e_lambda}
    \bm e(t+\Delta t) = & \ \bm e'(t+\Delta t) + \frac{b \Delta t^2}{2 I}\bm F'(t), \\  \label{eq:e_aux}
    \bm e'(t+\Delta t) = & \ \bm e(t) + b \Delta t \dot{\bm e} (t)  + \frac{{\gamma_r} b \dt}{2 I} \left(\Delta \bm \varGamma^e(t+\Delta t) \times \bm e(t)\right) ,\\ \label{eq:e_dot}
    \dot{\bm e}(t+ \Delta t) = & \ \dot{\bm e}(t) + \frac{\Delta t}{2 I}\left(\bm F'(t) + \bm F'(t+\Delta t)\right) 
      - \frac{\gamma_r}{I}\left( \bm e(t+\Delta t) - \bm e(t) \right)    + \frac{\gamma_r}{I}
     \left(\Delta \bm \varGamma^e(t+\Delta t) \times \bm e(t)\right) ,
\end{align}
\end{widetext}
with the constraint force
\begin{equation} \label{eq:force_constraint}
    \bm F'(t) = \lambda \bm e(t) ,
\end{equation}
$b=1/(1+\gamma_r \Delta t/(2I))$, and the Gaussian random number $\Delta \bm \varGamma^e(t)$ of zero mean and the second moments
\begin{equation}
    \lla \Delta \varGamma^e_{\alpha}(t) \Delta \varGamma^e_{\beta}(t+\Delta t) \rra = 2 D_r \delta_{\alpha \beta} \Delta t. 
\end{equation} 
The Lagrangian multiplier in Eq.~\eqref{eq:e_lambda} is determined by the constraint $ \bm e(t+\Delta t)^2 = 1$. In contrast to the SHAKE~\cite{cicc:86} or RATTLE~\cite{ande:83} algorithms, the Lagrangian multiplier is exactly calculated by insertion of Eq.~\eqref{eq:e_lambda} in the constraint equation, which yields a quadratic equation for $\lambda$ with the solution
\begin{equation}
\lambda = - \frac{2I}{b \dt^2} \left[1 - \sqrt{2 - \bm e'(t+\dt)^2} \right] ,
\end{equation}
where we assume $\bm e(t) \cdot \dot {\bm e}(t)=0$.
In case of $\bm e(t) = \bm e'(t)$ follows $\lambda =0$, as expected. Insertion of Eq.~\eqref{eq:e_aux} gives
\begin{equation} \label{eq:lagpar_fin}
\lambda = - \frac{2I}{b \dt^2} \left[1- \sqrt{1 -(b \Delta t)^2 \left(\dot{\bm e} + \frac{\gamma_r}{2I}  \Delta \bm \varGamma^e  \times \bm e \right)^2} \right] .
\end{equation}
Independent of $I$ and $\gamma_r$, the quadratic term in this equation is much smaller than zero as long as $D_r \Delta t \ll 1$, hence,
\begin{equation} \label{eq:lag_par_approx}
\lambda =  - b I \left[\dot{\bm e}(t) + \frac{\gamma_r}{2I}(\Delta \bm \varGamma^e (t+\Delta t) \times \bm e(t)) \right]^2 .
\end{equation}
In the asymptotic limits $I \to 0$ and $I \to \infty$, we then find for a fixed $\dt$
\begin{align} \label{app:lagpar_limit}
\lambda = \displaystyle 
\left\{
\begin{array}{cc} \displaystyle 
- \frac{\gamma_r}{2 \Delta t} (\Delta \bm \varGamma^e (t+\Delta t) \times \bm e(t))^2 \ , & I \to 0 \\[1ex] \displaystyle 
 - I \dot{\bm e}(t) ^2 \ , & I \to \infty
\end{array}
\right.
\end{align}
The average of the stochastic term is $\langle (\Delta \bm \varGamma^e \times \bm e)^2 \rangle =2 \gamma_R \Delta t =  2 (d-1) D_r \Delta t$, where $d$ is the spatial dimension. Thus, in the overdamped limit, Eq.~\eqref{eq:e_lambda} becomes
\begin{equation}
     \Delta \bm e = \Delta \bm \varGamma^e(t+\Delta t) \times \bm e(t)   -(d-1) D_r \bm e(t) \dt    .
\end{equation}
This is the Langevin equation within the Ito interpretation of the stochastic process.~\cite{raib:04,wink:15}

Equation~\eqref{eq:e_dot} does not exactly satisfy the condition
\begin{equation} \label{eq:const_e_dot}
    \bm e(t+\Delta t) \cdot \dot{\bm e}(t+\Delta t) =0 
\end{equation}  
following from the constraint $\bm e^2(t+\Delta t) =1$.
To ensure the condition, we adopt the RATTLE scheme~\cite{ande:83} and set $\bm F'(t+\dt) = \lambda^v \bm e(t+\dt)$.  Equation~\eqref{eq:e_dot} then reads
\begin{widetext}
\begin{equation}  \label{eq:e_dot_lag}
    \dot{\bm e}(t+\Delta t) = \dot{\bm e}(t) + \frac{\Delta t}{2 I}\left( \lambda \bm e(t) + \lambda^v \bm e(t+\Delta t)\right) 
      - \frac{\gamma_r}{I}\left( \bm e(t+\Delta t) - \bm e(t) \right)    + \frac{\gamma_r}{I}
     \left(\Delta \bm \varGamma^e(t+\Delta t) \times \bm e(t)\right) .
\end{equation}
\end{widetext}
Insertion of Eq.~\eqref{eq:e_dot_lag} into Eq.~\eqref{eq:const_e_dot} yields
\begin{widetext}
\begin{equation} \label{eq:lambda_u}
  \lambda^v = - \frac{2I}{\Delta t} \left\{ \dot{\bm e}(t)\cdot \bm e(t+\Delta t) + \frac{\Delta t}{2 I} \lambda \bm e(t) \cdot \bm e(t+\Delta t)  
      - \frac{\gamma_r}{I}\left[ 1 - \bm e(t)\cdot \bm e(t+\Delta t) \right]    + \frac{\gamma_r}{I}
     \left[\bm \varGamma^e(t+\Delta t) \times \bm e(t)\right] \cdot \bm e(t+\Delta t) \right\}.
\end{equation}
\end{widetext}
Here, $\lambda$ is the Lagrangian multiplier of Eq.~\eqref{eq:lagpar_fin}. Insertion of the constraint force $\bm F'(t+\Delta t) = \lambda^v \bm e(t+\dt)$ into Eq.~\eqref{eq:e_dot} gives the velocity $\dot{\bm e}(t+\dt)$.

\end{appendix}


\end{document}